\documentclass[twocolumn,showpacs,preprintnumbers,floatfix,amsmath,amssymb,prc]{revtex4}

\usepackage{dcolumn}
\usepackage{bm}
\usepackage{graphicx}
\usepackage{isotope}

\newcommand{\ve}[1]{\ensuremath{\mathbf{#1}}}
\newcommand{\n}[1]{\ensuremath{|\mathbf{#1}|}}
\newcommand{\EB}{\ensuremath{\epsilon_B}}
\newcommand{\nC}{\ensuremath{n_t^\text{corr}}}
\newcommand{\nMF}{\ensuremath{n_t^\text{MF}}}
\newcommand{\nTot}{\ensuremath{n_t}}
\newcommand{\PMF}{\ensuremath{P_t^\text{MF}}}
\newcommand{\PC}{\ensuremath{P_t^\text{corr}}}

\begin{document}
\title{Construction of spectral functions for medium-mass nuclei}

\author{Artur M. Ankowski}
\email{artank@ift.uni.wroc.pl}
\author{Jan T. Sobczyk}

\affiliation{Institute of Theoretical Physics\\ University of
Wroc{\l}aw,\\ pl. Maksa Borna 9, 50-204 Wroc{\l}aw, Poland}

\date{\today}

\graphicspath{{plot/}}

\begin{abstract}
This article is aimed at improving the description of lepton-nucleus interactions in the sub-GeV energy range. Approximate spectral functions for oxygen, calcium, and argon are constructed and used to obtain the electron cross sections in a~given scattering angle. Comparison with a~sample of available experimental data shows satisfactory agreement. Discrepancy between the presented model and the systematic computations available for oxygen [O.~Benhar {\it et al.}, Phys. Rev. D {\bf 72}, 053005 (2005)] is also found to be very small. Analysis of appropriate kinematical regions leads to the conclusion that the obtained argon spectral function should describe well neutrino scattering in the 800-MeV energy region. Several approximations used in the model are critically reviewed. All the details needed to implement the presented approach in Monte Carlo simulations are given.
\end{abstract}


\maketitle
\section{Introduction}\label{sec:Introduction}

In the impulse approximation (IA), lepton-nucleus interaction is described as a two step process: in the first step, the lepton interacts with a single bound nucleon, and in the second one, the resulting particles propagate inside the nucleus. The IA formalism is the basic framework in which $\sim$1-GeV leptons scattering is described: on one hand, it is the approach applied to understand electron scattering, and, on the other, it is used to model neutrino interactions~\cite{ref:NuInt}. The status of these two cases is quite different: for electrons a lot of experimental data exist, whereas for neutrinos precise measurements are still missing. Due to this lack of knowledge, reliable theoretical models are needed in the next generation of precise neutrino oscillation experiments~\cite{ref:NextGen}, possibly with liquid argon target.

To construct a successful model of neutrino-nucleus scattering, the following procedure seems to be well justified:
\begin{itemize}
\item[(i)]{relevant kinematical region in energy and momentum transfer has to be identified,}
\item[(ii)]{description of a~nucleus for electron scattering should be formulated in this kinematical region,}
\item[(iii)]{performance of the electron scattering model must be confronted with the existing data, and, if the agreement is satisfactory,}
\item[(iv)]{the same treatment of nuclear effects should be applied to neutrino interactions.}
\end{itemize}
This is the basic logic of this paper, in which we propose a~model to describe the $\sim$1-GeV neutrino scattering off medium-mass nuclei, such as calcium and argon. The article reports continuation of the research started in Ref.~\cite{ref:Ankowski&Sobczyk}, where a~less sophisticated description was applied to argon, and Ref.~\cite{ref:Ankowski}, where the model was introduced.

In the IA regime, a nucleus is described by means of the spectral function (SF). The SF contains information about the momentum distribution in conjunction with the distribution of binding energy of nucleons inside the the nucleus. Evaluation of the SF for medium nuclei requires several approximations. In our presentation, we try to identify and justify all the theoretical assumptions, but the most important argument for the correctness of our model is the agreement of its predictions with the data for electron scattering. Detailed verification of the description is performed using two targets, namely oxygen and calcium. Oxygen was selected because of additional opportunity to compare results with a~more systematic theoretical approach to modeling of the SF~\cite{ref:Benhar&Farina&Nakamura}, whereas the calcium nucleus is most similar to argon's, for which precise measurements have been performed. Our description of \isotope[40][20]{Ca} is confronted also with the theoretical results of Butkevich and Mikheyev~\cite{ref:Butkevich&Mikheyev}. Finally, a comparison with the few known data for electron scattering off  argon~\cite{ref:Anghinolfi_Ar} is done as well.

Basic computations of quasi{\-}elastic inclusive scattering are standard and can be found elsewhere~\cite{ref:Frullani&Mougey}. The outcome of numerical calculations depends on several assumptions
which specify the implementation of the SF model. We apply the recent BBBA05 parameterization of the proton and neutron form factors~\cite{ref:BBBA05}. The off-shell hadronic current matrix elements are evaluated with the use of the standard de Forest prescription~\cite{ref:deForest}. Furthermore, in the electromagnetic case, we adopt a~procedure to impose electromagnetic current conservation. Such prescription is not unique, and for this reason, in the case of
weak interactions, we avoid analogous manipulations with the vector part of the current and rely on the de Forest approach only.

An important ingredient in calculations is the treatment of final-state interactions (FSI). There are several approaches to deal with them, e.g., the Wentzel-Kramers-Brillouin method, relativistic mean-field approach~\cite{ref:Maieron&al}, or correlated Glauber approximation~\cite{ref:FSI_Benhar&al}. In the previous article~\cite{ref:Ankowski&Sobczyk}, we adopted the plane wave impulse approximation (PWIA) and disregarded FSI beyond the Pauli blocking, arguing that this approach is sufficient to describe neutrino scattering. In this paper, the presented model is validated by confronting it with a large sample of electron scattering data and this comparison requires inclusion of FSI. We consider two FSI effects: Pauli blocking and reinteractions of the struck nucleon with the spectator system described by means of the time-independent optical potential~\cite{ref:FSI_Benhar&al,ref:FSI_Nakamura&Seki&Sakuda,ref:FSI_Co'}.

In the lepton energy range of $\sim$1~GeV, two dynamical mechanisms are most important: quasielastic (QE) scattering (throughout this article we use the terminology of the neutrino community) and single pion production through the $\Delta$ excitation. They clearly manifest themselves as two peaks in the electron differential cross section in energy transfer for fixed scattering angle. Our numerical computations include the QE dynamics only. The reason is that our concern is to define a~systematic procedure to construct SF, and it is sufficient to test it in the case of the
QE process. Thank to this, we avoid dynamical issues in which the theoretical situation is not completely clear, namely the nonresonant background and two-particles--two-holes excitations~\cite{ref:Oset&al_nonres,ref:Oset&Salcedo_nonres}.
However, we have to pay the price for the constraint on the dynamics we adopted; in comparisons of our predictions with the experimental data, for higher values of the energy transfer some strength is systematically missing. Strictly speaking, we verify our model only in the kinematical region of energy transfers below the QE peak.

This paper is organized as follows. In Sec.~\ref{sec:Description}, the construction of our model is described: In Sec.~\ref{sec:General}, we present basic formulas for the lepton-nucleus cross section and introduce notation used throughout the article. Section~\ref{sec:Selection} discusses a~relation between kinematical regions in electron and neutrino scattering. In Sec.~\ref{sec:Model}, a~method to approximate SFs is given. The treatment of FSI effects in our model is covered in Sec.~\ref{sec:FSI}. Section~\ref{sec:Implementation} provides the parameterization of the SFs for oxygen, calcium, and argon. The information is detailed enough for everybody to be able to implement our results in their own numerical codes.

In Sec.~\ref{sec:Results}, our results are compared to large data samples for electron scattering off oxygen, calcium, and argon, selected according to the conclusions of Sec.~\ref{sec:Selection}. Our predictions are also confronted with other theoretical approaches. We observe that the performance of the presented description of nuclei is satisfactory and arrive at the conclusion that when applied to neutrino scattering, the model should produce reliable results.

Sec.~\ref{sec:Precision} is devoted to a~discussion of the approximations used in this paper. We consider plausible modifications of the adopted parameters and try to understand how uncertain our results for the cross section are.

Finally, in Sec.~\ref{sec:Summary}, we summarize the conclusions of this article. Since the most important features of the predictions seem to follow from the very basic assumption of the IA, the failure of the model in some kinematical situations may be interpreted as a failure of the IA itself. The presented results suggest that the IA starts to be unreliable when the typical value of momentum transfer is smaller than $\sim$350--400~MeV.

\section{Description of the model}\label{sec:Description}

\subsection{General information}\label{sec:General}

We consider quasielastic (QE) electron scattering off a~($Z$, $N$) nucleus of mass~$M_A$, which changes its four-momentum from $k\equiv(E_{\ve k},\ve k)$ to $k'\equiv(E_{\ve k'},\ve k')$. Associated with this interaction energy and momentum transfers are $\omega\equiv E_{\ve k}-E_{\ve k'}$ and $\ve q\equiv\ve k-\ve k'$, respectively. When the impulse approximation holds, i.e., when only one nucleon is involved in the primary vertex, nuclear effects can be described by means of the spectral function.

The proton spectral function (SF) of a~given nucleus $P_{(p)}(\ve p, E)$ is the probability distribution of removing from this nucleus a~proton with momentum $\ve p$ and leaving the residual nucleus with energy
\[
E_R=M_A-M+E+T_{A-1}~,
\]
which includes recoil energy of the residual nucleus $T_{A-1}=\ve p^2/(2M_{A-1})$, compare Refs.~\cite{ref:Frullani&Mougey,ref:Gross&Lipperheide}. The neutron SF is defined in an analogous way.

Energy balance of QE production of a~free nucleon carrying four-momentum $p'=(E_{\ve p'},\ve p')$,
\[
\omega+M_A=E_R+E_{\ve p'}~,
\]
may be rewritten in a~useful form using {\it removal} energy~$E$, which is an argument of the SF:
\[
\omega+M-E=T_{A-1}+E_{\ve p'}~.
\]
In Sec.~\ref{sec:Model}, we will justify that the recoil energy can be neglected, and therefore from now on, the energy balance
\begin{equation}\label{eq:energyBalance}
\omega+M-E=E_{\ve p'}~
\end{equation}
is used.

According to the IA, the inclusive electron-nucleus cross section is
the sum of contributions from protons and neutrons:
\[
\frac{d\sigma}{d\omega d\n q}=\frac{d\sigma_{(p)}}{d\omega d\n q}+\frac{d\sigma_{(n)}}{d\omega d\n q}~.
\]
Each term is expressed by the standard formula
\begin{eqnarray}\label{eq:crossSection}
\frac{d\sigma_t}{d\omega d\n q}&=&{2\pi\alpha^2}\frac{\n q}{E_{\ve k}^2}
\int dE\:d^3p\:\frac{P_t(\ve p, E)}{E_{\ve p}E_{\ve {p'}}}\\
& &\quad\times
\delta\big(\omega+M-E-E_\ve{p'}\big)L_{\mu\nu}^\text{em}H^{\mu\nu}_{\text{em, }t}\nonumber~,
\end{eqnarray}
where the index~$t$ denotes the nucleon isospin. The leptonic tensor is given by
\[
L_{\mu\nu}^\text{em}=2(k_\mu k'_\nu+k'_\mu k_\nu-k\cdot k'\thinspace g_{\mu\nu})~,
\]
due to negligible mass of electron, and the hadronic tensor is
\begin{eqnarray*}
H^{\mu\nu}_{\text{em, }t}&=&M^2 H_{1,\thinspace t}\Big(-g^{\mu\nu}+\frac{q^\mu q^\nu}{q^2}\Big)\\
& &+H_{2,\thinspace t}\Big(p^\mu-\frac{p\cdot q}{q^2}q^\mu\Big)\Big(p^\nu-\frac{p\cdot q}{q^2}q^\nu\Big)~,%
\end{eqnarray*}
with the scalar coefficients $H_{1,\thinspace t}$ and $H_{2,\thinspace t}$ depending on $q^2\equiv \omega^2-\n q^2$ and $\tau=-q^2/(4M^2)$ in the following way:
\begin{eqnarray*}
H_{1,\thinspace t}&=&\tau (F_{1,\thinspace t}+F_{2,\thinspace t})^2~,\\
H_{2,\thinspace t}&=&F_{1,\thinspace t}^2+\tau F_{2,\thinspace t}^2~.
\end{eqnarray*}
The form factors $F_{i,\thinspace t}=F_{i,\thinspace t}(q^2)$ are in turn expressed by the appropriate electric $G_{e,\thinspace t}$ and magnetic $G_{m,\thinspace t}$ form factors~\cite{ref:BBBA05}.

To handle the problem with the off-shell kinematics, we use the de Forest prescription~\cite{ref:deForest}: treat interacting nucleon as free and use free form factors but modify the energy conservation to take into account that a~part of energy transferred by the probe is absorbed by the spectator system. Comparing Eq.~\eqref{eq:energyBalance} to the energy balance
\[
\widetilde\omega+E_{\ve p}=E_{\ve p'}~,
\]
where the part of energy transfer which goes to the on-shell interacting nucleon with $p\equiv(E_{\ve p},\ve p)$ is denoted by~$\widetilde\omega$, one can find momentum-dependent binding energy:
\[
\EB=E_{\ve p}-M+E~.
\]
Replacing $q\equiv(\omega,\ve q)$ by $\widetilde q\equiv(\widetilde\omega,\ve q)=(\omega-\EB,\ve q)$ in the hadronic tensor,
\begin{equation}\label{eq:deForest}
H^{\mu\nu}_{\text{em, }t}\rightarrow \widetilde H^{\mu\nu}_{\text{em, }t}~,
\end{equation}
we obtain the standard description of the off-shell kinematics.

However, this procedure violates the conservation of the electromagnetic current, because $q_\mu\widetilde H^{\mu\nu}_{\text{em, }t}\neq 0$. To restore it, we have to add a~correction to the contraction of the tensors:
\begin{equation}\label{eq:CECRestoration}
L_{\mu\nu}^\text{em}\widetilde H^{\mu\nu}_{\text{em, }t}\rightarrow L_{\mu\nu}^\text{em}\widetilde H^{\mu\nu}_{\text{em, }t}+L_{\mu\nu}^\text{em}\widetilde H^{\mu\nu}_{\text{cor, }t}~,
\end{equation}
which is equal to
\begin{equation}
L_{\mu\nu}^\text{em}\widetilde H^{\mu\nu}_{\text{cor, }t}=\frac{M^2}{\widetilde q^2}c_1 \widetilde H_{1,\thinspace t}+c_2 \widetilde H_{2,\thinspace t}~.
\end{equation}
The coefficients~$c_1$ and $c_2$ can be expressed as
\begin{eqnarray*}
c_1&=&(\omega-\widetilde\omega)\Big[(\mathcal{Q}^2-\omega^2)(\omega+\widetilde\omega)\\
& &\qquad\qquad\qquad-4(\n k\n q\mathcal{Q}-\omega\ve k\cdot\ve q) \Big]~,\\
c_2&=&c_1\mathcal{P}^2+4(\omega-\widetilde\omega)\mathcal{PQ}\:{\left(\ve k\cdot\ve p-\frac{\ve p\cdot\ve q}{\ve q^2}\ve k\cdot\ve q\right)}~,\\
\end{eqnarray*}
with a~shorthand notation introduced for
\begin{eqnarray*}
\mathcal{Q}&=&\frac{2\ve k\cdot\ve q}{\n q}-\n q~,\\
\mathcal{P}&=&\frac1{2\n q}(2E_{\ve p}+\omega)~.\\
\end{eqnarray*}

When we consider QE muon neutrino scattering, its four-momentum is denoted as $k\equiv(E_{\nu},\ve k)$, four-momentum of the produced muon as $k'\equiv(E_{\mu},\ve k)$, and $q\equiv (\omega,\ve q)\equiv k-k'$. The cross section
\begin{eqnarray}\label{eq:crossSectionW}
\frac{d\sigma^\text{weak}}{d\omega d\n q}&=&\frac{G^2_F\cos^2\theta_C}{4\pi}\frac{\n q}{E_{\nu}^2}
\int dE\:d^3p\:\frac{P_{(n)}(\ve p, E)}{E_{\ve p}E_{\ve {p'}}}\\
& &\quad\times
\delta(\omega+M-E-E_\ve{p'})L_{\mu\nu}^\text{weak}H^{\mu\nu}_\text{weak}~\nonumber
\end{eqnarray}
contains contraction of the leptonic and hadronic tensors
\begin{eqnarray*}
L_{\mu\nu}^\text{weak}&=&2(k_\mu k'_\nu+k'_\mu k_\nu-k\cdot k'\thinspace g_{\mu\nu}-i\epsilon_{\mu\nu\rho\sigma}k^\rho k'^\sigma),\\%
H^{\mu\nu}_\text{weak}&=&-g^{\mu\nu}M^2H_1+p^\mu p^\nu H_2+\frac{i}2\varepsilon^{\mu\nu\kappa\lambda}p_\kappa q_\lambda H_3\nonumber\\&&-q^\mu q^\nu H_4+\frac12(p^\mu q^\nu +q^\mu p^\nu)H_5,
\end{eqnarray*}
where
\begin{eqnarray*}
H_1&=&F_A^2(1+\tau)+\tau (F_1+F_2)^2,\\
H_2&=&F_A^2+F_1^2+\tau F_2^2,\\
H_3&=&2 F_A(F_1+F_2),\\
H_4&=&\frac14 F_2^2(1-\tau)+\frac12 F_1 F_2+F_A F_P-\tau F_P^2\\
H_5&=&H_2.
\end{eqnarray*}
The tensors differ from the ones for electromagnetic interaction due to the  axial contribution (in our calculations axial mass $M_A=1.03$~GeV) and to the fact that, thanks to the conserved-vector-current hypothesis, $F_1$ and $F_2$ are expressed by differences of the proton and neutron form factors; see~Ref.~\cite{ref:BBBA05}. Considering neutrino interactions, we apply the de Forest prescription~\eqref{eq:deForest} but do not restore conservation of the vector current. All the other quantities are defined and denoted as in the case of electron interaction.

\subsection{Selection of the electron data}\label{sec:Selection}

\begin{figure}
        \includegraphics[width=0.46\textwidth]{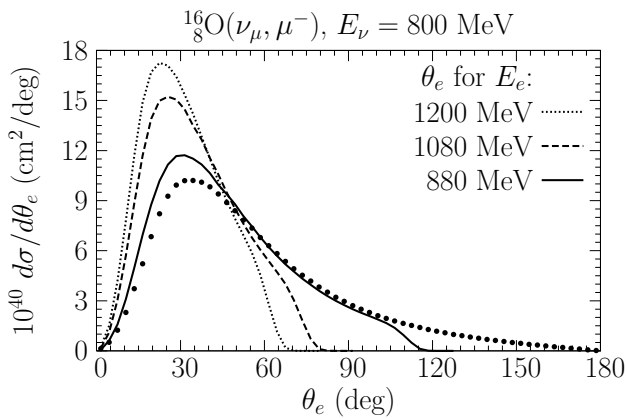}%
\caption{\label{fig:mapping} Analysis of the dependence of the
$\isotope[16]{O}(\nu_\mu,\mu^-)$ cross section on energy- and
momentum-transfer. Lines show what electron scattering
angles~$\theta_e$ in the process $\isotope[16]{O}(e,e')$ correspond
to the same kinematical region. The standard
differential cross section in muon scattering angle for
$\isotope[16]{O}(\nu_\mu,\mu^-)$ is represented by points.}
\end{figure}

According to the plan outlined in Sec.~\ref{sec:Introduction}, first we identify the region in the $(\omega, \n q)$ plane which is most important for QE \emph{neutrino} scattering. The energy and momentum transfers are related to the muon production angle~$\theta$ by the expression
\begin{equation}\label{eq:omega,q_Neutrino}
\cos\theta=\frac{E_\nu-\omega}{\n{k'}}+\frac{\omega^2-\ve q^2-m_\mu^2}{2E_\nu\n{k'}}~,%
\end{equation}
where $\n{k'}=\sqrt{(E_\nu-\omega)^2-m_\mu^2}$. Therefore fixing~$\theta$ is equivalent to restricting a~region in the $(\omega, \n q)$ plane. Points in Fig.~\ref{fig:mapping} show the neutrino differential cross section~$d\sigma^\text{weak}/d\theta$ for neutrino energy $E_\nu=0.8$ GeV. The peak at $\sim$33$^\circ$ is rather broad and $\sim$50\% of the cross section comes from $\theta\in[20^\circ;~56^\circ]$. For $E_\nu=1.2$ GeV, the maximum moves to $\sim$22$^\circ$~and the peak becomes narrower (not shown in the figure).

We want to map the allowed kinematical region for neutrino scattering, weighted by the cross section, to the corresponding region for \emph{electron} scattering. For electron of energy~$E_e$, the relation analogous to Eq.~\eqref{eq:omega,q_Neutrino} reads
\begin{equation}\label{eq:omega,q_Electron}
\cos\theta_e=1+\frac{\omega^2-\ve q^2}{2E_e(E_e-\omega)}~.%
\end{equation}
Hence, for a~given value of $E_\nu$ and selected $E_e$, we can map the muon production angle to the electron scattering angle:
\[
\theta\mapsto\theta_e~.%
\]
To weight the electron scattering angles by the neutrino cross section, we calculated the quantity
\begin{equation}\label{eq:mapper}
\frac{d\theta}{d\theta_e}\frac{d\sigma^\text{weak}}{d\theta}~.%
\end{equation}
using the oxygen target, described by the Benhar SF. (We checked that the Fermi gas model and the effective description~\cite{ref:Ankowski&Sobczyk} lead to the same conclusions.)

From now on, we concentrate on $E_\nu=0.8$ MeV. In Fig.~\ref{fig:mapping}, we show the quantity~\eqref{eq:mapper} for three selected values of electron energy: $E_e=0.88$, 1.08, and 1.2~GeV. The conclusion is that to describe well the 0.8-GeV neutrino scattering, our model should be verified with 1.2-GeV electron data at $\theta_e\sim23^\circ$, $1.08$-GeV data at $\theta_e\sim25^\circ$, or $0.88$-GeV data at $\theta_e\sim30^\circ$.

Let us go into more detail. We deduce that for $E_e=1.2$ GeV, the range of muon scattering angle $[20^\circ;~56^\circ]$ corresponds to $\theta_e\in[15^\circ;~36^\circ]$ with a~maximum at 23$^\circ$; for $E_e=1.08$ GeV, to $\theta_e\in[17^\circ;~39^\circ]$ with a~maximum at 25$^\circ$; whereas for $E_e=0.88$ GeV, to $\theta_e\in[19^\circ;~50^\circ]$ with a~maximum at 30$^\circ$. The general rule is that for higher
electron beam energies, the smaller scattering angles become significant.

Equation~\eqref{eq:omega,q_Electron} is well defined when $E_e\geq E_\nu$. For lower $E_e$, this equation may be applied only for the prize of a~loss of normalization---the form of the denominator excludes some of the points in the $(\omega, \n q)$ plane. For example, when $E_e=0.73$ GeV is used, 5\% of the strength is lost and $\theta\in[20^\circ;~56^\circ]$ corresponds to $\theta_e\in[22^\circ;~61^\circ]$ with a~maximum at 35$^\circ$.

In the case of electron scattering off oxygen, the measurements were performed for scattering angle~$32^\circ$ using beam energies 700, 880, 1080, 1200, and 1500~MeV~\cite{ref:Anghinolfi,ref:Anghinolfi_Ar}, whereas 537- and 730-MeV beams were used for angle~$37.1^\circ$~\cite{ref:O'Connell}. As follows from our analysis, to obtain the model which describes well QE neutrino-nucleus scattering at energy 800~MeV, the most significant electron data are those for 880 and 730~MeV. The relevance of the experimental points for 1080 and 700~MeV is smaller. The energy and momentum transfers which characterize scattering with electron beams of energies 1200 and 537~MeV are
least similar to what is needed but these energies are still in the region of interest. The set of data for 1500~MeV was collected at too high scattering angle for our applications.

Among a few papers reporting results of electron scattering experiments with a~calcium
target~\cite{ref:Whitney,ref:Meziani,ref:Yates,ref:Williamson}, the most suitable for testing of our model is
Ref.~\cite{ref:Williamson}, containing data at the lowest scattering angle, namely 45.5$^\circ$, in conjunction with the highest values of beam energy---up to 841~MeV. We have checked that all the measurements at 45.5$^\circ$ correspond to our region of interest in the $(\omega, \n q)$ plane. Obviously, only the data for $E_e=841$~MeV cover the whole region, and the lower electron energy is, the more normalization is lost. For example, when one uses $E_e=545$ MeV, $\theta\in[20^\circ;~56^\circ]$ corresponds to $\theta_e\in[29^\circ;~75^\circ]$ with a~maximum at 46$^\circ$ and 27.4\% of the strength is lost. Therefore, we rely mainly on comparisons with the experimental
data for higher electron energies.

Finally we want to explain why we decided to study neutrino energy $E_\nu=0.8$~GeV. The reason is that there is a lot of relevant electron scattering data to compare with. For higher $E_\nu$, say 1.2~GeV, the situation would be quite different---the electron-scattering data at smaller angles would be required, but they are missing for the targets we are interested in.

\subsection{How we model spectral function}\label{sec:Model}

The spectral function describes distribution of nucleons in the $(\ve p, E)$ plane. By integrating out the dependence on $E$, the momentum distribution~$\nTot(\ve p)$ is obtained:
\begin{equation}\label{eq:n(p)_def}
\nTot(\ve p)=\int P_t(\ve p, E)\,d E~.
\end{equation}
Our normalization convention is
\begin{equation}
\int P_t(\ve p, E)\,d^3p\,d E=N_t~,
\end{equation}
where the number of nucleons $N_t$ is $Z$ for protons and $N$ for neutrons.

Approximately 80\%--90\% of nucleons in a~nucleus can be described as occupying shell-model states and moving freely in the mean-field (MF) potential. The rest of them take part in interactions. It is natural to decompose the SF into the sum of the MF and correlated parts~\cite{ref:Ciofi&Simula&Frankfurt&Strikman,ref:Kulagin&Petti,ref:Benhar&Fabrocini&Fantoni&Sick}:
\begin{equation}\label{eq:PMF+PC}
P_t(\ve p, E)=N_t\left[\PMF(\ve p,E)+\PC(\ve p,E)\right]~.%
\end{equation}
By analogy to Eq.~\eqref{eq:n(p)_def}, the MF and correlated momentum distributions are introduced:
\begin{eqnarray}
\nMF(\ve p)&=&\int\PMF(\ve p,E)\,d E~,\label{eq:nMF_def}\\
\nC(\ve p)&=&\int\PC(\ve p,E)\,d E~,\label{eq:nCorr_def}
\end{eqnarray}
so the momentum distribution can be described as composed of two subdistributions:
\begin{equation}\label{eq:nMF+nC}
\nTot(\ve p)=N_t\left[\nMF(\ve p)+\nC(\ve p)\right]~.%
\end{equation}

\subsubsection{Treatment of the MF part}\label{sec:MF}%

The basic assumption underlying the presented approach is the IA, therefore the MF part of the SF can be written in the form (compare~\cite{ref:Ciofi&Simula&Frankfurt&Strikman,ref:Kulagin&Petti,ref:Benhar&Farina&Nakamura}):
\begin{equation}\label{eq:PMF_def}
\PMF(\ve p,E)=\sum_{\alpha}\frac{c_\alpha}{N_t}\: |\phi_\alpha(\ve p)|^2F_\alpha\big(E_{\alpha}+T_{A-1}-E\big)~,%
\end{equation}
with separated contributions from each shell-model state~$\alpha$, $\alpha$ ranging from 1 to~$N_t$. Denoting spectroscopic factor by $c_\alpha$, wave function by $\phi_\alpha(\ve p)$, level energy by $E_{\alpha}$, and a function describing level width by $F_\alpha$, we have omitted the isospin index~$t$ for clarity of the notation. If interactions between nucleons disappeared, the MF part would describe the whole SF (equivalently, all $c_\alpha$'s would become equal to 1) and each $F_\alpha$ would be the $\delta$ function.

In this article we are interested in a description of medium-sized nuclei, like calcium and argon. Recoil energy of the residual nucleus $T_{A-1}$ may then be neglected in the MF part of the SF since it is typically $\sim$0.5 MeV (see the average MF momenta in Table~\ref{tab:MomDistrib}).

We assume that $\int F_\alpha(E)\:dE=1$, what can be physically interpreted as the momentum independence of level widths. Than the MF momentum distribution~\eqref{eq:nMF_def} can be expressed as
\begin{equation}
\nMF(\ve p)=\sum_{\alpha} \frac{c_\alpha}{N_t}|\phi_\alpha(\ve p)|^2~.%
\end{equation}
Let us make the \emph{crucial assumption}: each level contributes equally to the MF momentum distribution. It means that in Eq.~\eqref{eq:PMF_def} for each $\alpha$ we can make the substitution
\begin{equation}\label{eq:eachLevSameMD}
c_\alpha|\phi_\alpha(\ve p)|^2\rightarrow\nMF(\ve p).
\end{equation}
The final form of the MF part of the SF,
\begin{equation}\label{eq:PMF}
\PMF(\ve p,E)=\nMF(\ve p)\frac1{N_t}\sum_{\alpha} F_\alpha(E_{\alpha}-E)~,%
\end{equation}
have to be further specified by the form of the function which describes level width. For a given half-width, the Breit-Wigner distribution has longer tails then the Gaussian one, so we found the latter more suitable:
\begin{equation}\label{eq:Gauss}
F_\alpha(x)=\sqrt{\frac{8}{\pi D_\alpha^2}}\exp\left(-8{x^2}/{D_\alpha^2}\right)~.%
\end{equation}
Therefore we refer to the proposed model as the \emph{Gaussian spectral function} (GSF). The factor 8 in the argument of exponential function is introduced for further convenience.

Note that the sum in Eq.~\eqref{eq:PMF} extends to \emph{all} occupied states. This approach differs from the one presented in Ref.~\cite{ref:Ankowski&Sobczyk} and allows to avoid singularities in the argon SF.

To describe a~specific nucleus by its Gaussian SF, one needs to know the appropriate MF momentum distribution, the values of energy levels, and their widths~$D_\alpha$.

\subsubsection{Approach to the correlated part}\label{sec:Corr}%

Interacting nucleons are described by the correlated part of the SF. It is a~known fact (see Ref.~\cite{ref:Ciofi&Liuti&Simula} and references therein) that the two-nucleon interactions dominate.
These short-range correlations give rise to pairs of nucleons with high relative momentum. We follow the approach of
Kulagin and Petti~\cite{ref:Kulagin&Petti} and do not include in the considerations interactions of higher order. Than, \PC~can be expressed analytically in the form:
\begin{eqnarray}\label{eq:correlationSF}
\PC(\ve p,E)&=&\nC(\ve p)\frac{M}{\n p}\sqrt{\frac\alpha\pi}\nonumber\\%
& &\times\left[\exp(-\alpha \ve p_\text{min}^2)-\exp(-\alpha\ve p_\text{max}^2)\right]~.\qquad%
\end{eqnarray}
The constant~$\alpha$ appearing in the above formula is a~shorthand notation for $3/(4\langle \ve p_\text{MF}^2\rangle \beta)$ with $\beta=(A-2)/(A-1)$ and the mean
square of the MF momentum $\langle \ve p_\text{MF}^2\rangle$ defined as
\begin{equation}\label{eq:<p^2_MF>}
\langle \ve p_\text{MF}^2\rangle=\frac{\int\ve p^2\nMF(\ve p) d^3p}{\int\nMF(\ve p) d^3p}~,
\end{equation}
whereas
\begin{equation}\label{eq:p_minR,p_maxR}\begin{split}
{\ve p}_\text{min}^2&=\Big\{\beta \n p - \sqrt{2M\beta[E-E^{(2)}-T_{A-1}]}\,\Big\}^2,\\%
{\ve p}_\text{max}^2&=\Big\{\beta \n p + \sqrt{2M\beta[E-E^{(2)}-T_{A-1}]}\,\Big\}^2.%
\end{split}\end{equation}
The two-nucleon separation energy $E^{(2)}$ is an average excitation of the $(A-2)$ nucleon system. Since by definition averaging should be carried out only over the low-lying states, it can be approximated by the mass difference $E^{(2)}=M_{A-2}+2M-M_A$.

Because an overwhelming contribution to the correlated part comes from the peak at
\[
E\approx E^{(2)}+\frac{\ve p^2}{2M}
\]
and the recoil energy~$T_{A-1}$ is less than $\ve p^2/(2M)$ by the factor~$(A-1)$, therefore Eq.~\eqref{eq:p_minR,p_maxR} may be simplified to
\begin{equation}\label{eq:p_min,p_max}\begin{split}
{\ve p}_\text{min}^2&=\Big\{\beta \n p - \sqrt{2M\beta[E-E^{(2)}]}\,\Big\}^2,\\%
{\ve p}_\text{max}^2&=\Big\{\beta \n p + \sqrt{2M\beta[E-E^{(2)}]}\,\Big\}^2.%
\end{split}\end{equation}
For the lightest considered here nucleus, i.e., oxygen this simplification yields a~$\lesssim 0.2\%$ change of the cross section.

\subsection{How we apply FSI}\label{sec:FSI}

The struck nucleon moves in nuclear matter and may interact with surrounding spectators. Such interactions make the nucleon an open system in the sense that measured $E_\ve{p'}$ is not equal to its energy in the interaction vertex. One can describe this situation in terms of a~complex optical potential, $U=V-iW$, as proposed originally in Ref.~\cite{ref:FSI_Horikawa&al.}. We assume that the potential is time-independent. Then the result is equivalent to making in Eq.~\eqref{eq:crossSection} the substitution
\begin{equation}\label{eq:FSI}
\delta(\dots)\rightarrow\frac{W/\pi}{W^2+[\dots-V]^2}~,
\end{equation}
see Refs.~\cite{ref:FSI_Benhar&al,ref:FSI_Nakamura&Seki&Sakuda}. The imaginary part of the optical potential may be approximated by
\begin{equation}\label{eq:imaginaryOP}
W=\frac{\hbar c}{2}\rho_\text{nucl}\sigma_{N\!N}\frac{\n{p'}}{E_\ve{p'}}~.
\end{equation}
This article's main interest is a description of medium nuclei, such as calcium and argon, therefore the nuclear matter density $\rho_\text{nucl}$ is assumed to be constant and equal to the saturation density $\rho_\text{sat}=0.16$~fm$^{-3}$. In the kinematical region of our interest, the typical proton kinetic energy is 100--300 MeV and the nucleon-nucleon cross section $\sigma_{N\!N}=\frac12(\sigma_{pp}+\sigma_{pn})$ at $\rho_\text{sat}$ varies between 16.2 and 19.1~mb~\cite{ref:Pandharipande&Pieper}. We set it to the~value for 200-MeV protons, i.e., to 17.4~mb.

\begin{figure}[t]
        \includegraphics[width=0.46\textwidth]{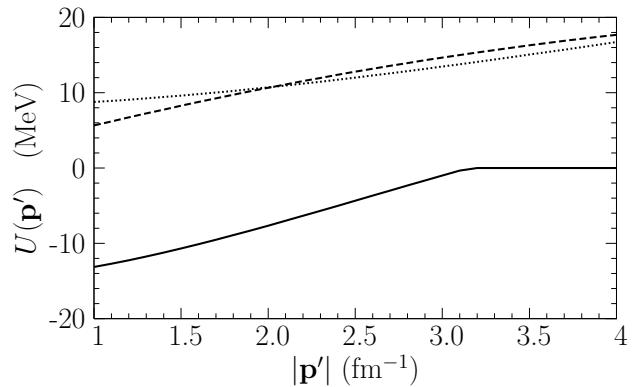}%
\caption{\label{fig:pot} Optical potential used in this paper. Dashed line represents its imaginary part obtained from Eq.~\eqref{eq:imaginaryOP} and solid line the real one from Eq.~\eqref{eq:OP}. For comparison, the imaginary part calculated also from Eq.~\eqref{eq:OP} is shown by dotted line.}
\end{figure}
The real part of the potential we use is calculated in the following way: Reference~\cite{ref:Cooper&al} gives a~Dirac optical potential of~\isotope[40][20]{Ca} fitted to proton-scattering data in the energy range 161--1040 MeV as a~function of kinetic energy of proton and position in the nucleus. Since what we need is the potential depending on energy only, averaging over spatial coordinate should be performed. We do it by evaluating the potential at the root mean square (rms) radii from Ref.~\cite{ref:Cooper&Hama&Clark&Mercer}. As a~result, we obtain a~potential $U(\ve{p'})$ related to the scalar and vector part of the potential in Ref.~\cite{ref:Cooper&al} by
\begin{equation}\label{eq:OP}
E_\ve{p'}+U(\ve{p'})=\sqrt{[M+S(T_\ve{p'},\bar r_S)]^2+\ve{p'}^2}+V(T_\ve{p'},\bar r_V)~.
\end{equation}
In the above equation, $\bar r_S$ denotes two parameters, because the real and imaginary part of $S$ have different values of the rms radius. The same holds true for $\bar r_V$ and $V$. For $\n{p'}>3.1$~fm$^{-1}$, the real part of $U(\ve{p'})$ is positive, what is inconsistent with the correlated Glauber theory~\cite{ref:Benhar&Fabrocini&Fantoni&Sick}. Therefore when $\n{p'}>3.1$~fm$^{-1}$, we set its value to zero, as shown in Fig.~\ref{fig:pot}.

From the few parameterizations of the potential in Ref.~\cite{ref:Cooper&al} we decided to use the one called case~2.
We checked that the imaginary part of $U(\ve{p'})$ is then very close to $W$ obtained from Eq.~\eqref{eq:imaginaryOP} (compare the dotted and dashed lines in Fig.~\ref{fig:pot}), so our approach is self-consistent.

The assumption that the optical potential is time-independent leads to folding of the cross section with the Lorentzian function [Eq.~\eqref{eq:FSI}]. To cure the resulting problem with nonzero cross section for~$\omega<0$ (compare Fig.~4 in Ref.~\cite{ref:FSI_Nakamura&Seki&Sakuda}), we impose an additional constraint on the upper limit of the integration over~$E$,
\[
E<\omega~.
\]
When FSI effects are not included, this restriction comes automatically from the energy-conserving $\delta$~function.

\begin{figure}[t]
        \includegraphics[width=0.46\textwidth]{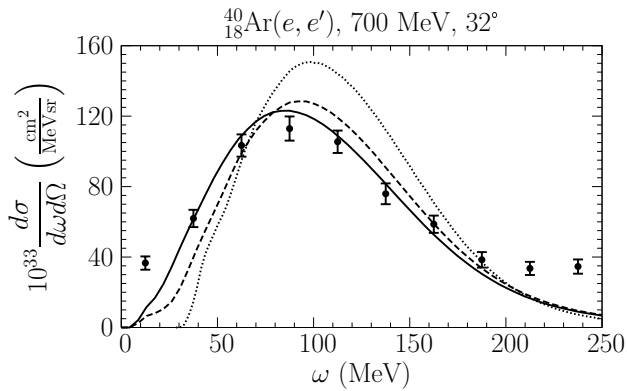}%
\caption{\label{fig:FSI} Influence of FSI on electron-nucleus cross section. Dotted line shows the cross section without FSI, dashed line with only imaginary part applied, and solid line with full FSI.}
\end{figure}

As can be seen in Fig.~\ref{fig:FSI}, the essential effect of the imaginary part of the potential is to broaden the QE peak, whereas the real part mainly moves the strength to lower~$\omega$'s. Thanks to these two effects, the agreement of the calculated cross sections with the experimental data is significantly better. However, the time-independent imaginary part of the optical potential overestimates the FSI, namely too much strength is redistributed from the QE peak to its tails. We postpone the discussion of this point to Sec.~\ref{sec:Precision}.

\subsection{Details of implementation}\label{sec:Implementation}

In this subsection, we want to cover all the details of description of three nuclei---oxygen, calcium, and argon---by their Gaussian SFs. A~procedure to divide momentum distributions given in Ref.~\cite{ref:Bisconti&Arias&Co} into the MF and correlated parts is presented and justified. We concentrate on $\nC(\ve p)$ and obtain the MF part from~\eqref{eq:nMF+nC}. Then, we show the parameterization of the energy levels and comment the way, their widths are obtained.

\subsubsection{Momentum distributions}

In Ref.~\cite{ref:Bisconti&Arias&Co,ref:Co'_priv}, the total momentum distributions for many nuclei are calculated. However, the model described in this paper requires a~separation of the MF and correlated contributions [see Eqs.~\eqref{eq:PMF} and~\eqref{eq:correlationSF}]. References~\cite{ref:MD_Benhar&al,ref:Ciofi&Simula} contain plots with momentum distributions divided in the way we need. The conclusion from these articles is that above $\n p=2$~fm$^{-1}$, the correlated part dominates overwhelmingly. We assume that above 2~fm$^{-1}$ this contribution is equal to the total distribution, and that \nC(\ve p) may be expressed as the correlated distributions given there, i.e., as a sum of two exponential functions. Moreover, smooth transition at 2~fm$^{-1}$ is imposed.

Distributions from Refs.~\cite{ref:Bisconti&Arias&Co,ref:Co'_priv}, denoted here by $n(t,\ve p)$, are calculated up to $\n p=3.585$~fm$^{-1}$. We extrapolated them smoothly to 5~fm$^{-1}$, but it turned out to have very little influence on the cross sections.

The correlated part of the momentum distribution is assumed to be of
the following functional form:
\begin{widetext}
\begin{equation}\label{eq:nCorrParametrization}
\nC(\ve p)=\begin{cases}
\frac{\mathcal{F}}{(2\pi)^3}\frac{A}{N_t}\left[C_1\exp(-e_1\ve p^2)+C_2\exp(-e_2\ve p^2)\right]&\text{for $0\leq\n p\leq2.025$~fm$^{-1}$,}\\[5pt]%
\frac{\mathcal{F}}{(2\pi)^3}\frac{A}{N_t}n(t,\ve p)&\text{for $2.025$~fm$^{-1}<\n p\leq3.585$~fm$^{-1}$,}\\[5pt]
\frac{\mathcal{F}}{(2\pi)^3}\frac{A}{N_t}C_3\exp(-e_3\ve
p^2)&\text{for 3.585~fm$^{-1}<\n p\leq5.0$~fm$^{-1}$.}
\end{cases}
\end{equation}
\end{widetext}
In the above equation, $A$ stands for the number of nucleons. We normalize the momentum distributions introducing the factor $\mathcal F$:
\[
\frac{\mathcal{F}}{(2\pi)^3}\frac{A}{N_t}\int_0^\text{5 fm$^{-1}$}4\pi \ve p^2n(t,\ve p)\:d\n p=1~.
\]
To find the values of the parameters in~\eqref{eq:nCorrParametrization}, we assume that $e_1\gg e_2$, so that only the $e_2$-containing term is responsible for the behavior of $\nC(\ve p)$ at large momenta. By demanding the continuity and smoothness of \nC, one can determine $C_2$ and $e_2$ at $\n p=2.025$ fm$^{-1}$, while $C_3$ and $e_3$ at $\n p=3.585$ fm$^{-1}$. The values of $e_1$ are taken from~\cite{ref:Ciofi&Simula}; in Sec.~\ref{sec:Precision} we will show that $e_1$'s do not affect the cross sections. The values of $C_1$ are fixed by the overall normalization of \nC, which follows from Ref.~\cite{ref:Bisconti&Arias&Co}: the data contained there in Tables~II and III allow to calculate what fraction of nucleons cannot be assigned to any shell-model state and, as a~consequence, must be described by the correlated part of SF. The normalization of \nC~with respect to~$\nTot$ is the same as the normalization of the the correlated SF with respect to the total SF.

\begin{figure}[t]
        \includegraphics[width=0.46\textwidth]{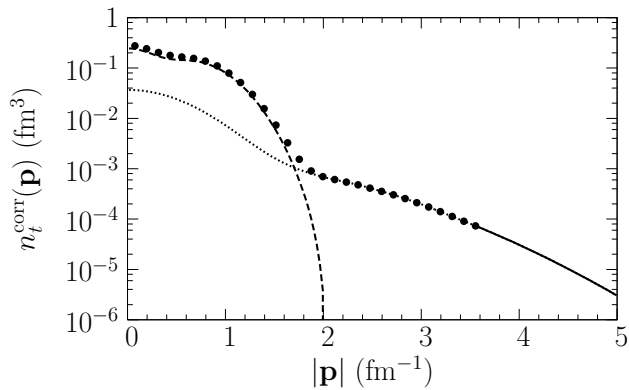}%
\caption{\label{fig:momCa40P} Proton momentum distribution in
\isotope[40][20]{Ca} from Ref.~\cite{ref:Bisconti&Arias&Co} (dots)
divided into the MF (dashed line) and correlated part (dotted and
solid line). Solid line shows extrapolation according to
Eq.~\eqref{eq:nCorrParametrization} with parameters given in
Table~\ref{tab:MomDistrib}.}
\end{figure}

\begin{table}
\caption{\label{tab:MomDistrib} Parameters of the correlated part~[see Eq.~\eqref{eq:nCorrParametrization}] of the momentum distributions from Ref.~\cite{ref:Bisconti&Arias&Co} and their normalization with respect to the total momentum distribution for various nuclei. The last row contains resulting values of the average mean field momentum defined in~\eqref{eq:<p^2_MF>}.}
\begin{ruledtabular}
\begin{tabular}{cccccc}
 & \multicolumn{1}{c}{\mbox{$\isotope[16][8]{O}$}} & \multicolumn{2}{c}{\mbox{$\isotope[40][20]{Ca}$}} & \multicolumn{2}{c}{\mbox{$\isotope[48][20]{Ca}$}}\\%
 & & \multicolumn{1}{r}{\mbox{proton}} & \multicolumn{1}{r}{\mbox{neutron}} & \multicolumn{1}{r}{\mbox{proton}} & \multicolumn{1}{r}{\mbox{neutron}}\\
 \hline
$\mathcal{F}$ & 1.0200 & 1.0370 & 1.0370 & 1.0440 & 1.0200 \\
$C_1$ & 2.1280 & 4.2150 & 4.2700 & 4.0040 & 4.6700 \\
$e_1$ & 1.4000 & 1.7700 & 1.7700 & 1.7700 & 1.7700 \\
$C_2$ & 0.1427 & 0.1940 & 0.1855 & 0.1536 & 0.1656 \\
$e_2$ & 0.2260 & 0.2260 & 0.2142 & 0.2018 & 0.2065 \\
$C_3$ & 0.1678 & 0.2282 & 0.2451 & 0.2500 & 0.2960 \\
$e_3$ & 0.2410 & 0.2580 & 0.2648 & 0.2972 & 0.2940 \\
Normal.&12.00\%&16.20\% & 16.20\% & 17.10\% & 13.64\% \\
\hline\\[-8pt]
$\sqrt{\langle \ve p^2_\text{MF}\rangle}$ (MeV)& 174.4 & 189.1 & 187.1 & 180.8 & 196.4\\
\end{tabular}
\end{ruledtabular}
\end{table}

Sample outcome of the described procedure is presented in Fig.~\ref{fig:momCa40P}. One can see that at the sewing points, $\n p=2.025$ fm$^{-1}$ and 3.585 fm$^{-1}$, the correlated contribution is smooth. The total momentum distributions and the normalization of the correlated parts alike are taken from Ref.~\cite{ref:Bisconti&Arias&Co}. Therefore the set of parameters collected in Table~\ref{tab:MomDistrib} can be considered as self-consistent.

To handle the lack of knowledge of the momentum distributions for protons and neutrons in the argon nucleus, we apply in the SFs the appropriate distributions calculated for \isotope[40][20]{Ca}.

\subsubsection{Description of the energy levels}

In our approach, each shell-model state~$\alpha$ is fully characterized by two
parameters: energy level~$E_\alpha$ and width~$D_\alpha$, defined by means of Eq.~\eqref{eq:Gauss}.

\begin{table}
\caption{\label{tab:CaEnergyLev} Energy
levels~$E_\alpha$~\cite{ref:CaPLev_Tornow&Chen&Delaroche,ref:CaNLev_Johnson&Mahaux} and widths~$D_\alpha$ for \isotope[40][20]{Ca}.}
\begin{ruledtabular}
\begin{tabular}{l|dd|dd}
 & \multicolumn{2}{d}{\mspace{25mu}\text{protons}} & \multicolumn{2}{d}{\mspace{25mu}\text{neutrons}}\\
 & \multicolumn{1}{c}{$E_\alpha$} & {D_\alpha} & \multicolumn{1}{c}{$E_\alpha$} & {D_\alpha}\\
 \hline
$1s_{1/2}$ & 57.38 & 25\footnotemark[1] & 66.12 & 25\footnotemark[1]\\
$1p_{3/2}$ & 36.52 & 15\footnotemark[1] & 43.80 & 15\footnotemark[1]\\
$1p_{1/2}$ & 31.62 & 15\footnotemark[1] & 39.12 & 15\footnotemark[1]\\
$1d_{5/2}$ & 14.95 & 4\footnotemark[1]  & 22.48 & 6\footnotemark[1]\\
$2s_{1/2}$ & 10.67 & 2\footnotemark[2]  & 17.53 & 4\footnotemark[2]\\
$1d_{3/2}$ & 8.88  & 2\footnotemark[2]  & 15.79 & 4\footnotemark[2]\\
$\alpha_F$ & 4.71  &     & 12.0 &  \\
\end{tabular}
\end{ruledtabular}
\footnotetext[1]{Fit to the plots from Refs.~\cite{ref:CaNLev_Johnson&Mahaux,ref:CaPLev_Tornow&Chen&Delaroche}.} \footnotetext[2]{Our estimate, details in text.}
\end{table}
\begin{table}
\caption{\label{tab:ArEnergyLev} Same as Table~\ref{tab:CaEnergyLev}, but for \isotope[40][18]{Ar}. Details in text.}
\begin{ruledtabular}
\begin{tabular}{l|dd|dd}
 & \multicolumn{2}{d}{\mspace{25mu}\text{protons}} & \multicolumn{2}{d}{\mspace{25mu}\text{neutrons}}\\
 & \multicolumn{1}{c}{$E_\alpha$} & {D_\alpha} & \multicolumn{1}{c}{$E_\alpha$} & {D_\alpha}\\
 \hline
$1s_{1/2}$ & 52\footnotemark[2] & 25 & 62    & 25\\
$1p_{3/2}$ & 32\footnotemark[2] & 15 & 40    & 15\\
$1p_{1/2}$ & 28\footnotemark[2] & 15 & 35    & 15\\
$1d_{5/2}$ & 11\footnotemark[2] & 4  & 18    & 5\\
$2s_{1/2}$ & 8\footnotemark[2] & 2  & 13.15\footnotemark[1] & 4\\
$1d_{3/2}$ & 6\footnotemark[2]  & 2  & 11.45\footnotemark[1] & 3\\
$1f_{7/2}$ &    &    & 5.56\footnotemark[1]  & 3 \\
$\alpha_F$ &    &    & 8.0\footnotemark[1] &  \\
\end{tabular}
\end{ruledtabular}
\footnotetext[1]{Theoretical calculations in~\cite{ref:ArLev_Johnson&Carlton&Winters}.}
\footnotetext[2]{Modified theoretical values for \isotope[40][20]Ca from~\cite{ref:CaPLev_Tornow&Chen&Delaroche}.}
\end{table}
\begin{table}
\caption{\label{tab:OxEnergyLev} Same as Table~\ref{tab:CaEnergyLev},
but for \isotope[16][8]{O}. The values of $D_\alpha$ are obtained differently; see details in text.}
\begin{ruledtabular}
\begin{tabular}{l|dd|dd}
 & \multicolumn{2}{d}{\mspace{25mu}\text{protons}} & \multicolumn{2}{d}{\mspace{25mu}\text{neutrons}}\\
 & \multicolumn{1}{c}{$E_\alpha$} & {D_\alpha} & \multicolumn{1}{c}{$E_\alpha$} & {D_\alpha}\\
 \hline
$1s_{1/2}$ & 45.00 & 70 & 47.00 & 70\\
$1p_{3/2}$ & 18.44 & 4 & 21.80 & 4\\
$1p_{1/2}$ & 12.11 & 4 & 15.65 & 4\\
\end{tabular}
\end{ruledtabular}
\end{table}

\begin{figure*}
    \begin{minipage}[l]{0.325\textwidth}
        \flushright
        \includegraphics[width=5.79cm]{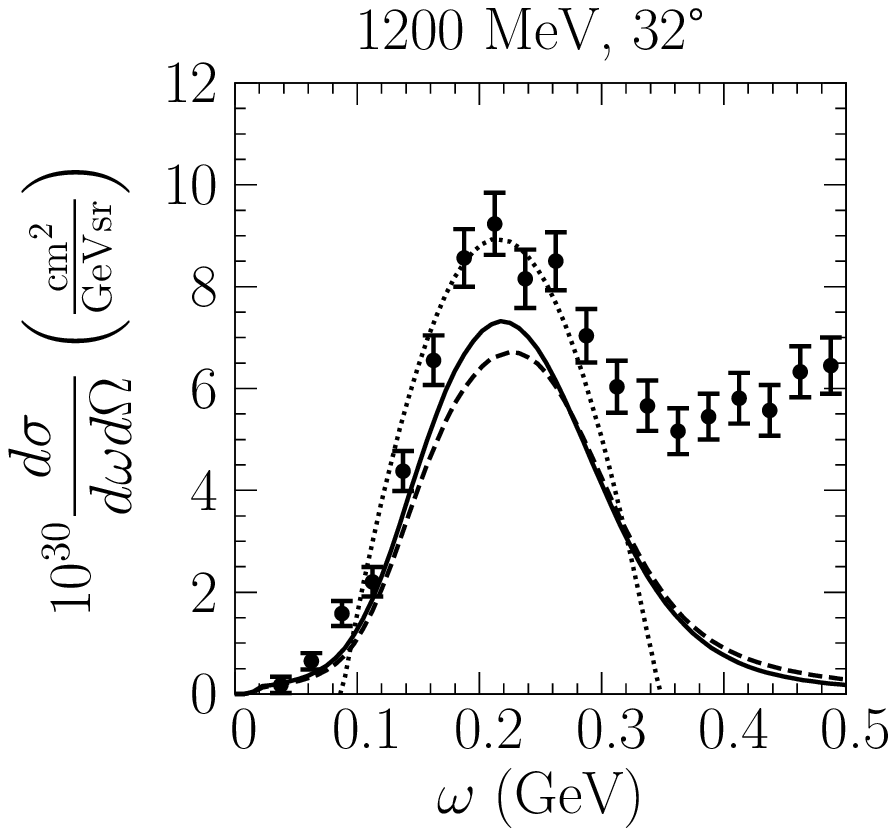}%
        \vspace{0.5cm}
    \end{minipage}
    \begin{minipage}[c]{0.325\textwidth}
        \flushright
        \includegraphics[width=4.8cm]{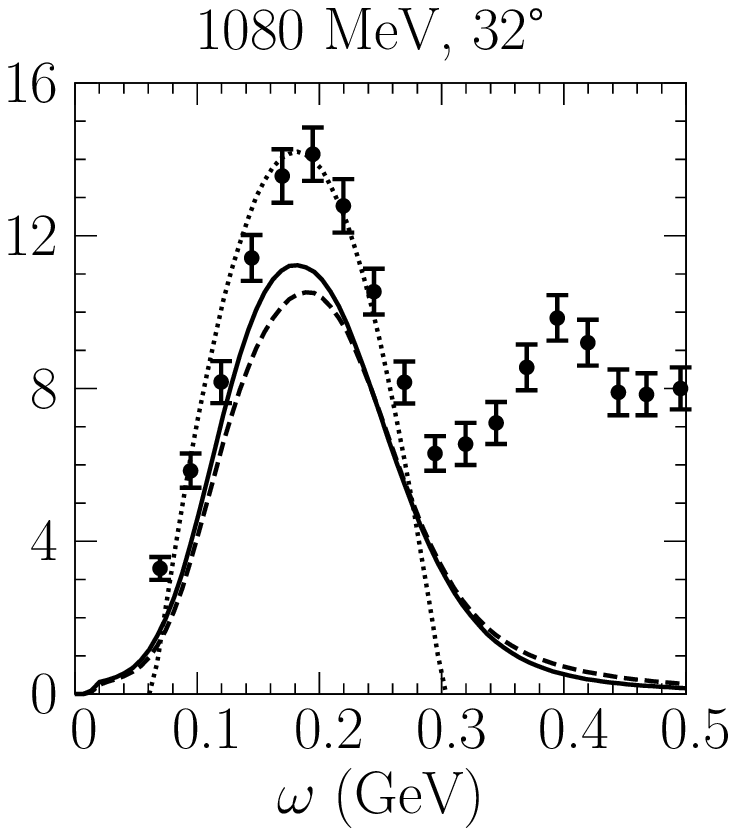}%
        \vspace{0.5cm}
    \end{minipage}
    \begin{minipage}[r]{0.325\textwidth}
        \flushright
        \includegraphics[width=4.8cm]{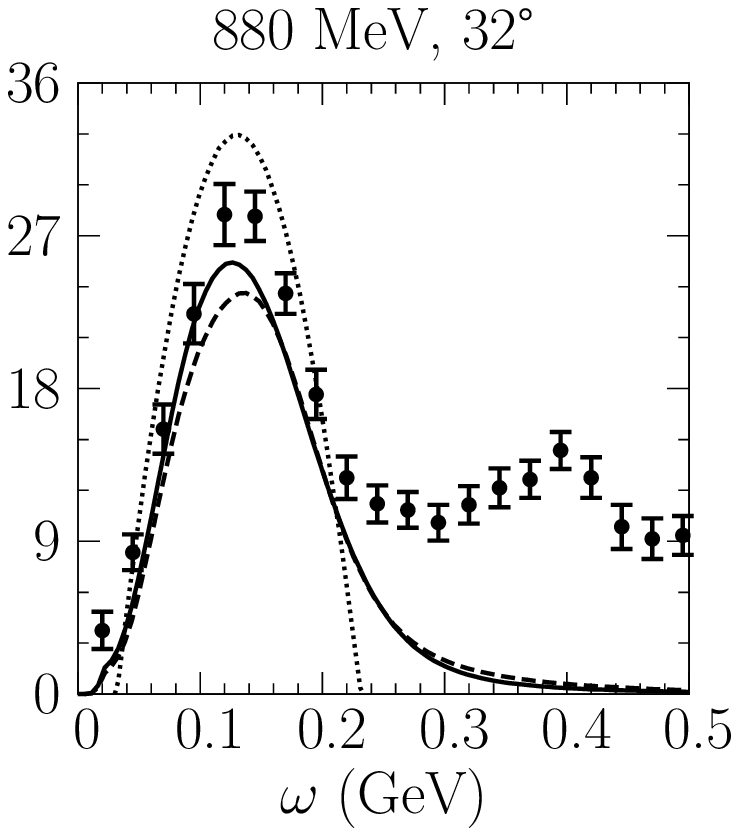}%
        \vspace{0.5cm}
    \end{minipage}
    %
    \begin{minipage}[l]{0.325\textwidth}
        \flushright
        \includegraphics[width=5.8cm]{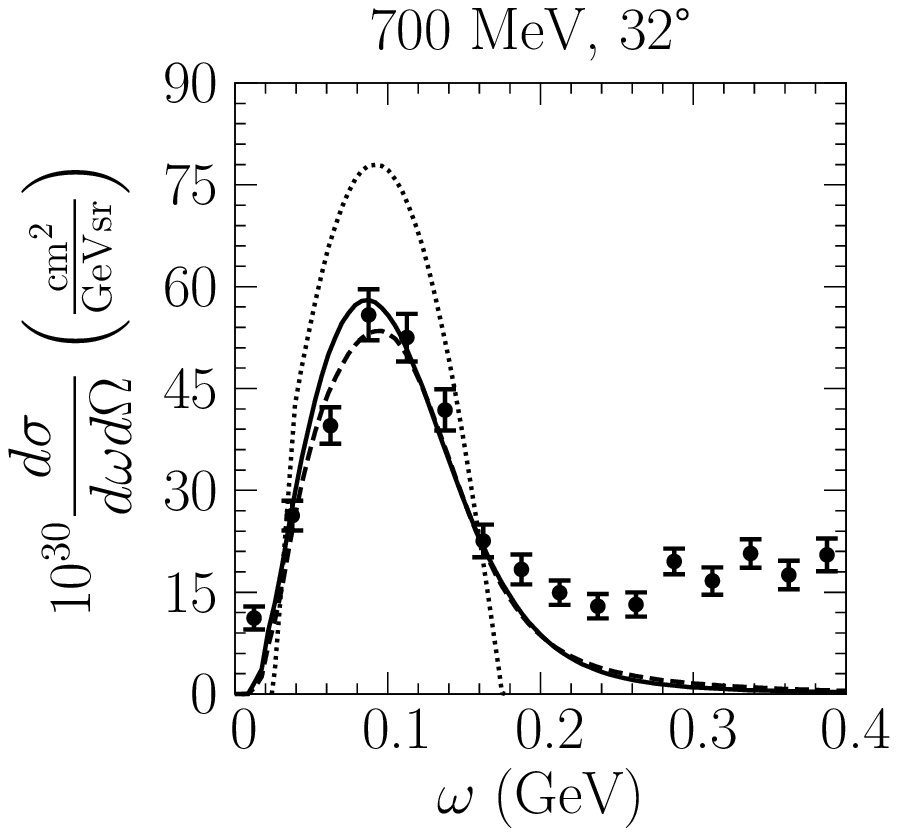}%
    \end{minipage}
    \begin{minipage}[c]{0.325\textwidth}
        \flushright
        \includegraphics[width=4.8cm]{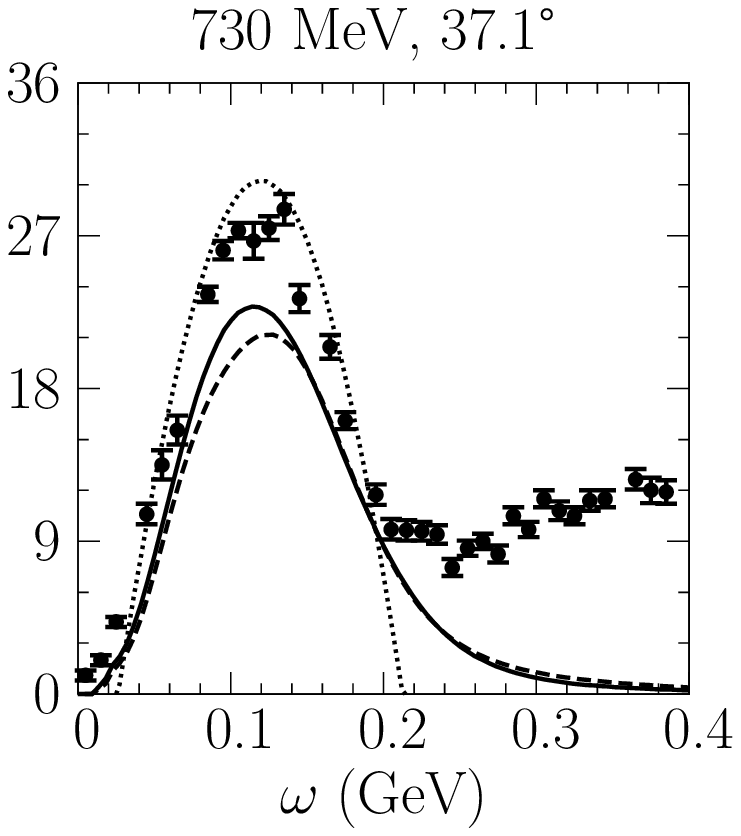}%
    \end{minipage}
    \begin{minipage}[r]{0.325\textwidth}
        \flushright
        \includegraphics[width=4.8cm]{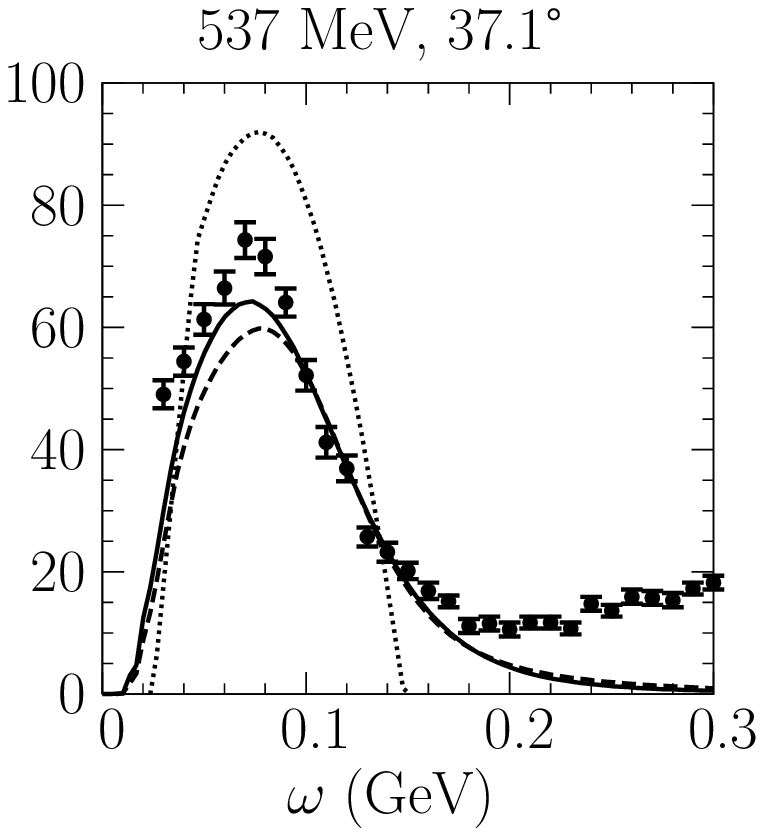}%
    \end{minipage}
\caption{\label{fig:Oxy} Cross sections of the process $\isotope[16]{O}(e,e')$ at miscellaneous values of beam energy for scattering angles $32^\circ$~\cite{ref:Anghinolfi,ref:Anghinolfi_Ar} and $37.1^\circ$~\cite{ref:O'Connell}. Results for the GSF (solid line) are compared to the Benhar SF~\cite{ref:Benhar&Farina&Nakamura} with the same FSI (dashed line) and the Fermi gas model without FSI (dotted line). The values of $\n q$ at the peaks are 637~MeV (for beam energy 1200~MeV), 573~MeV (for 1080~MeV), 466~MeV (for 880~MeV), 371~MeV (for 700~MeV), 441~MeV (for 730~MeV), and 325~MeV (for 537~MeV).}
\end{figure*}

\begin{figure*}
    \begin{minipage}[l]{0.325\textwidth}
        \flushright
        \includegraphics[width=5.79cm]{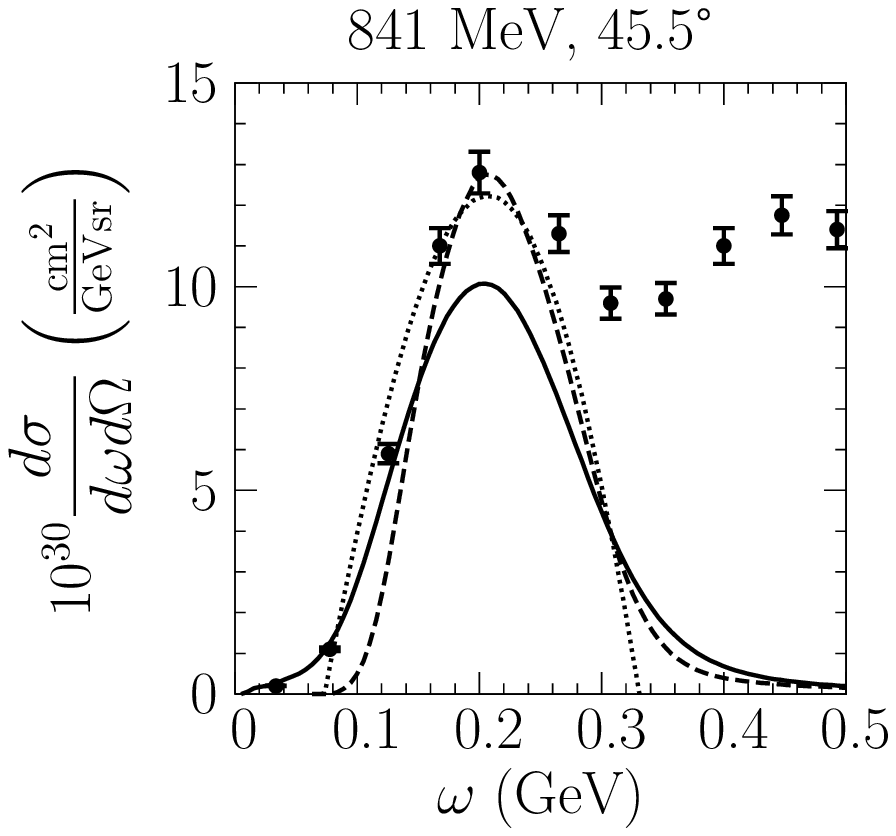}%
        \vspace{0.5cm}
    \end{minipage}
    \begin{minipage}[c]{0.325\textwidth}
        \flushright
        \includegraphics[width=4.8cm]{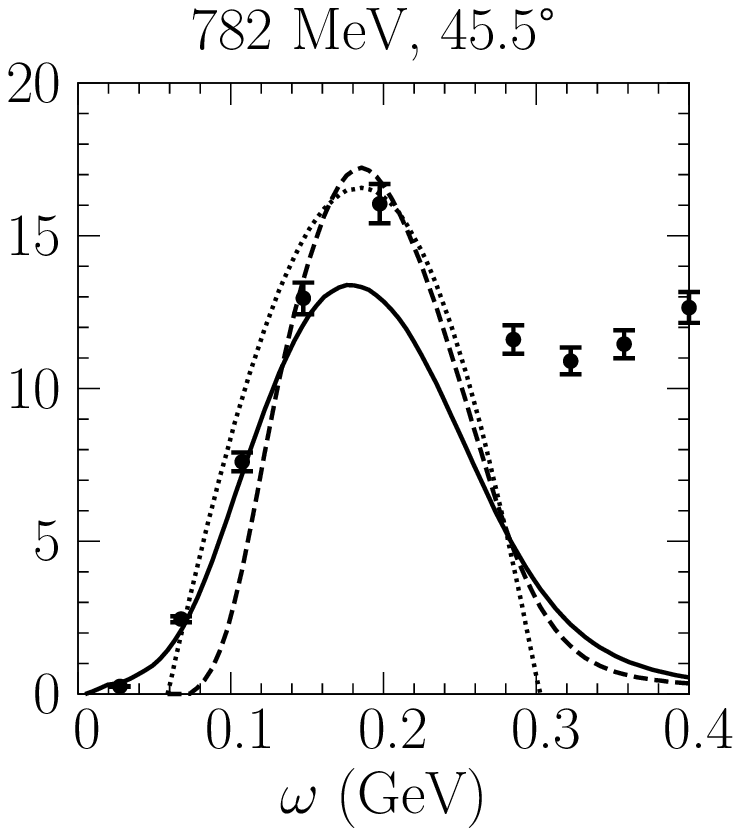}%
        \vspace{0.5cm}
    \end{minipage}
    \begin{minipage}[r]{0.325\textwidth}
        \flushright
        \includegraphics[width=4.8cm]{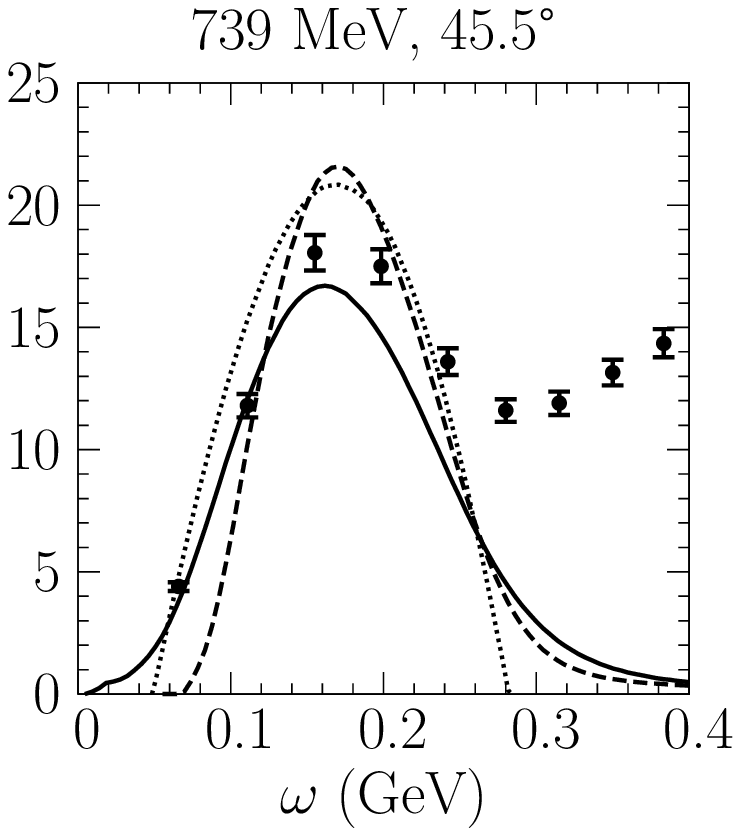}%
        \vspace{0.5cm}
    \end{minipage}
    %
    \begin{minipage}[l]{0.325\textwidth}
        \flushright
        \includegraphics[width=5.8cm]{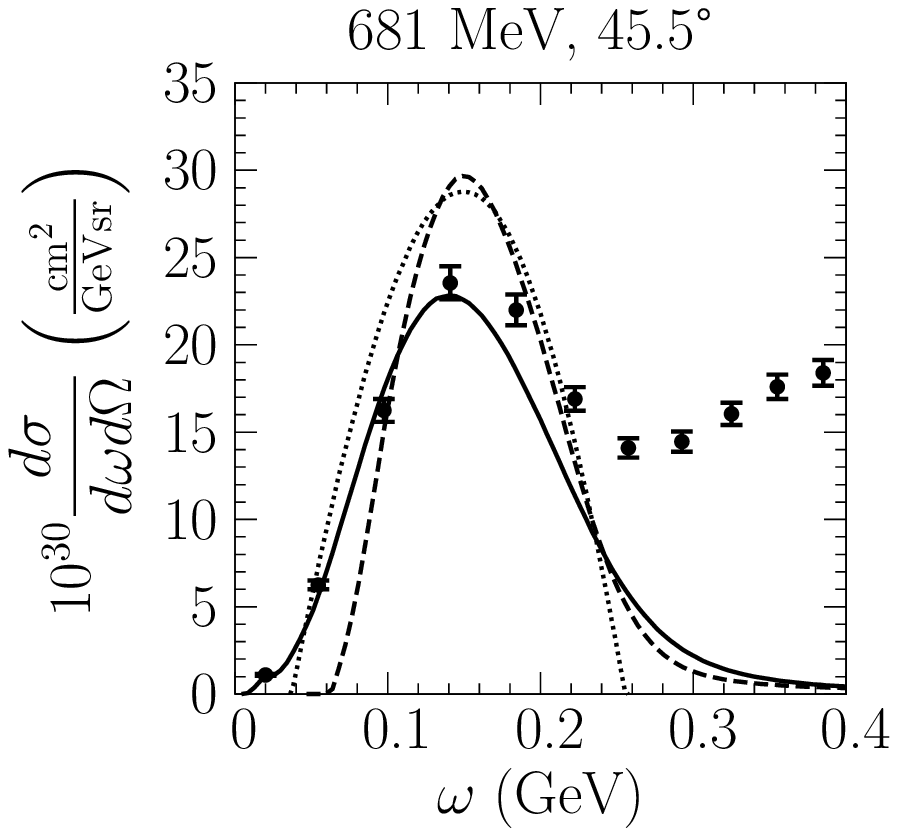}%
        \vspace{0.5cm}
    \end{minipage}
    \begin{minipage}[c]{0.325\textwidth}
        \flushright
        \includegraphics[width=4.8cm]{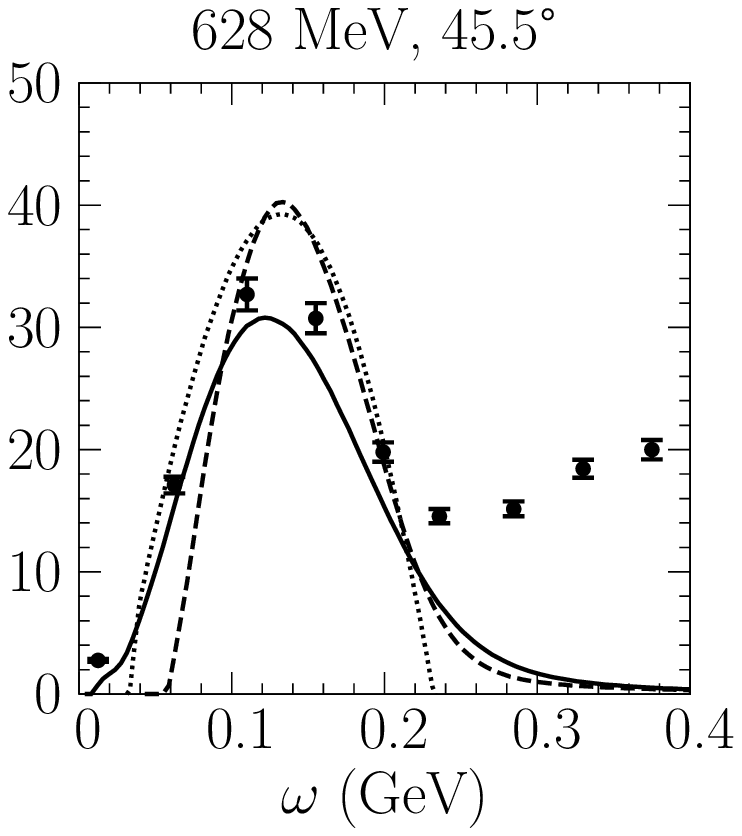}%
        \vspace{0.5cm}
    \end{minipage}
    \begin{minipage}[r]{0.325\textwidth}
        \flushright
        \includegraphics[width=4.8cm]{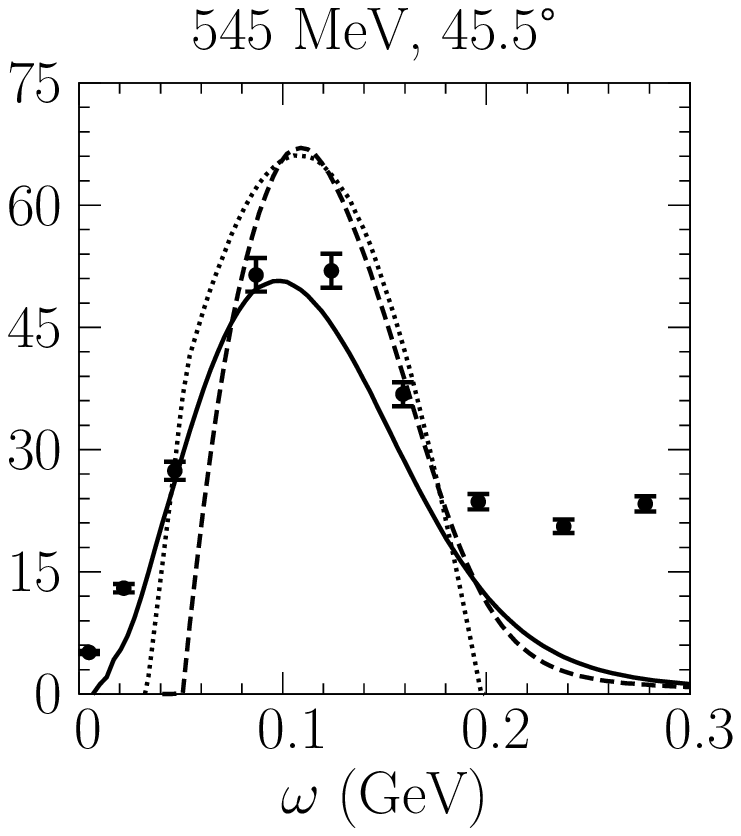}%
        \vspace{0.5cm}
    \end{minipage}
    %
    \begin{minipage}[l]{0.325\textwidth}
        \flushright
        \includegraphics[width=5.8cm]{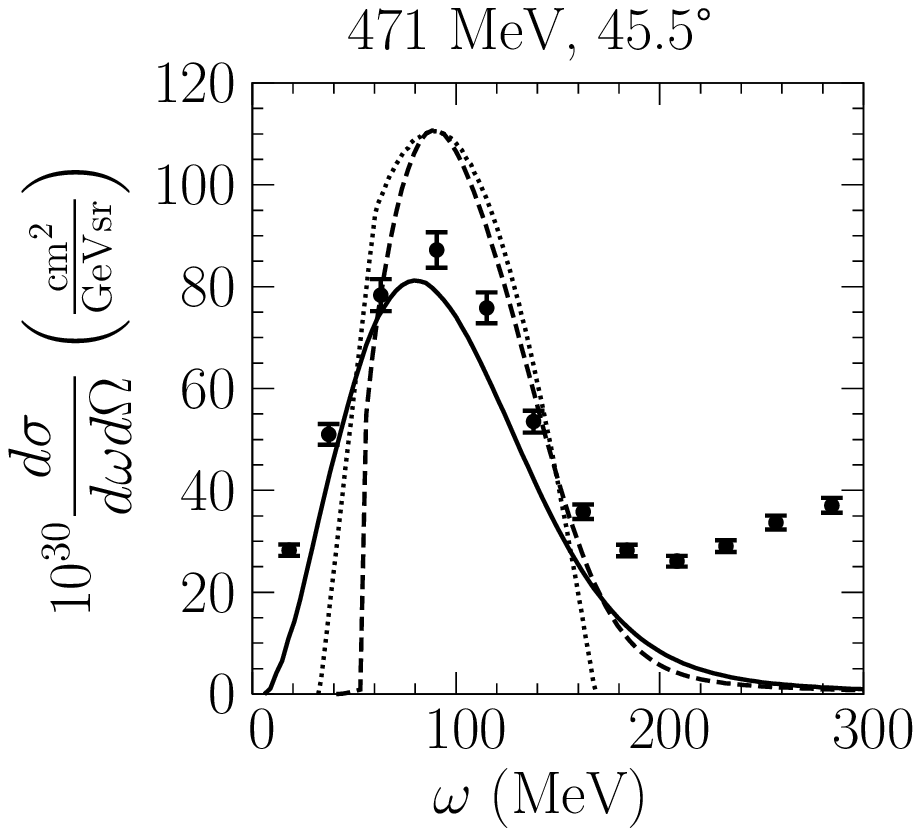}%
    \end{minipage}
    \begin{minipage}[c]{0.325\textwidth}
        \flushright
        \includegraphics[width=4.8cm]{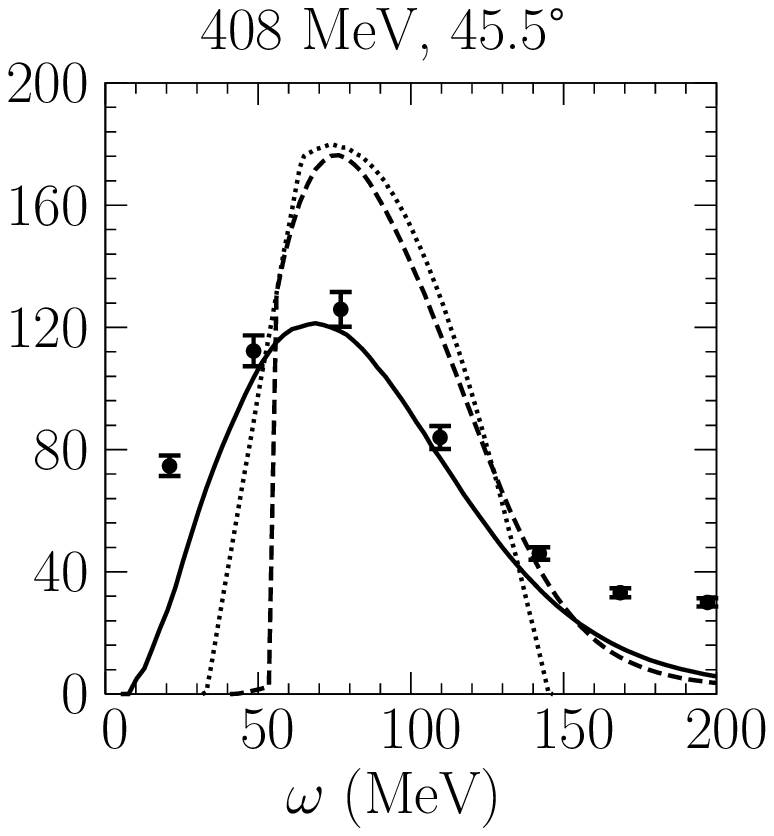}%
    \end{minipage}
    \begin{minipage}[r]{0.325\textwidth}
        \flushright
        \includegraphics[width=4.8cm]{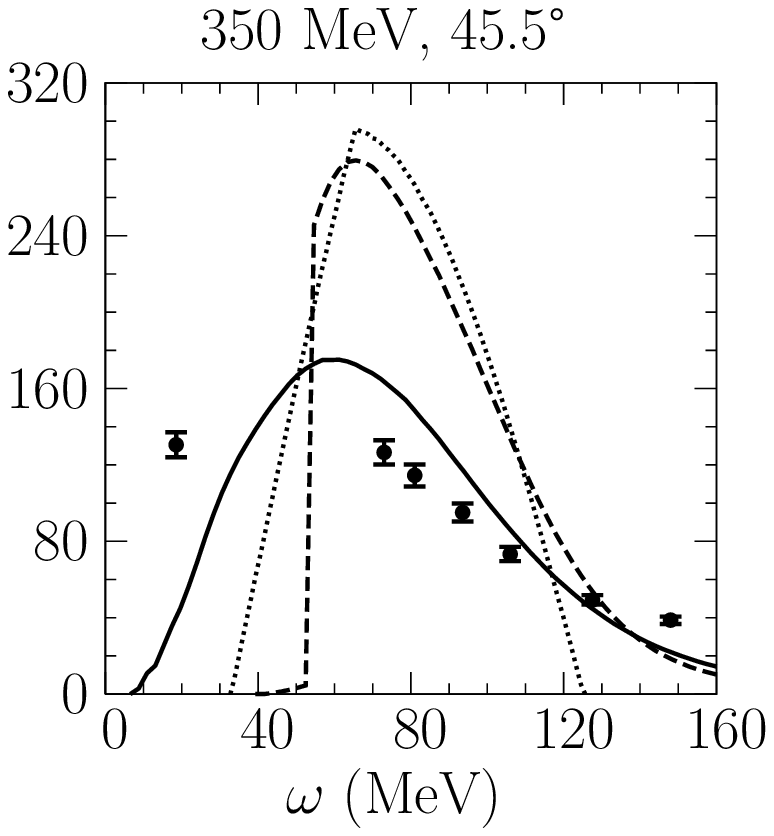}%
    \end{minipage}
\caption{\label{fig:Ca} Cross sections of $\isotope[40]{Ca}(e,e')$
scattering at angle $45.5^\circ$ and miscellaneous values of
electron beam energy~\cite{ref:Williamson}. Calculations for the GSF
(solid line) are compared to the results of Butkevich and
Mikheyev~\cite{ref:Butkevich&Mikheyev} (dashed line), and the Fermi
gas model (dotted line). The corresponding values of~$\n q$ at the
peaks are 602~MeV (for beam energy 841~MeV), 561~MeV (for 782~MeV),
531~MeV (for 739~MeV), 490~MeV (for 681~MeV), 453~MeV (for 628~MeV),
395~MeV (for 545~MeV), 342~MeV (for 471~MeV), 297~MeV (for 408~MeV),
and 254~MeV (for 350~MeV).}
\end{figure*}

\begin{figure*}
    \begin{minipage}[l]{0.48\textwidth}
        \flushleft
        \includegraphics[width=0.95\textwidth]{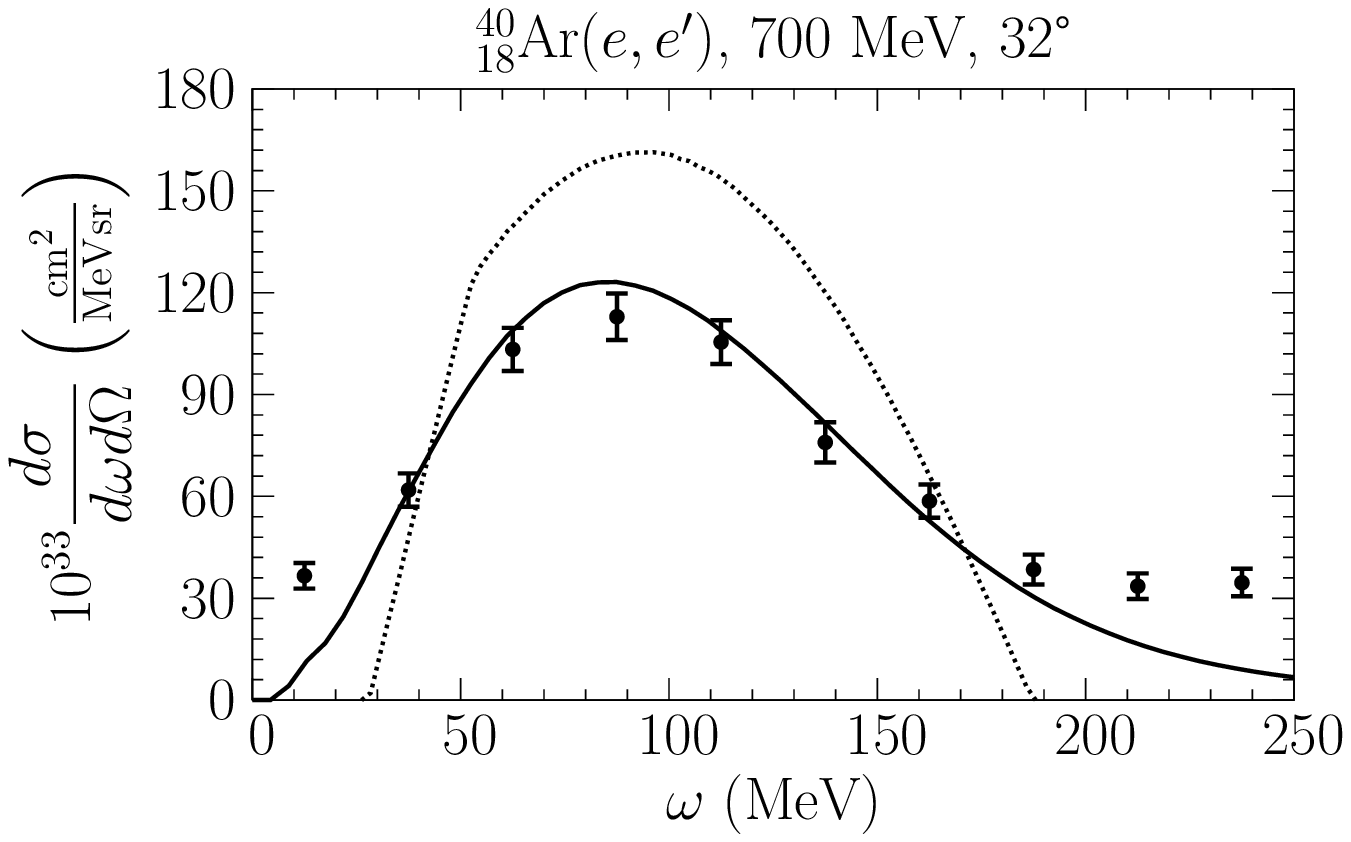}%
    \end{minipage}
    \begin{minipage}[r]{0.48\textwidth}
        \flushright
        \includegraphics[width=0.95\textwidth]{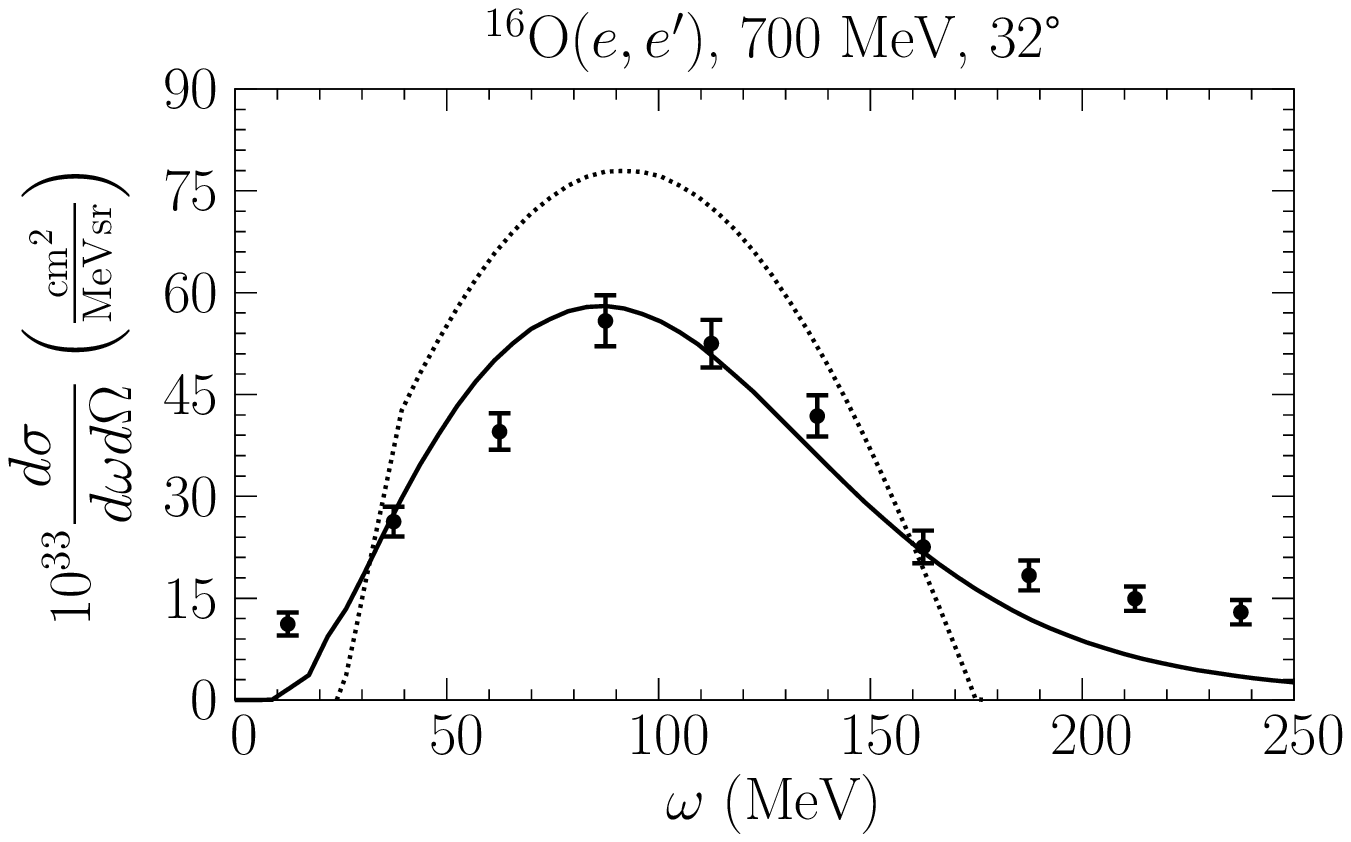}%
    \end{minipage}
\caption{\label{fig:Ar} Left panel: Comparison of the cross section
of  GSF (solid line) and the FG model (dotted line) with
experimental points for $\isotope{Ar}(e,e')$ at beam energy 700~MeV
and scattering angle $32^\circ$~\cite{ref:Anghinolfi_Ar}. Right
panel: Same, but for oxygen. Note that in both cases the similar
accuracy is obtained. The value of momentum transfer at the peaks
is 371~MeV.}
\end{figure*}

The energy levels of calcium shown in Table~\ref{tab:CaEnergyLev} result from theoretical calculations in Refs.~\cite{ref:CaNLev_Johnson&Mahaux} (for neutrons) and \cite{ref:CaPLev_Tornow&Chen&Delaroche} (for protons).
A~few available neutron levels of argon~\cite{ref:ArLev_Johnson&Carlton&Winters} form a pattern very similar to the one of the neutron levels of calcium: the distance between $1d_{3/2}$ and $2s_{1/2}$ is 1.7~MeV for Ar and 1.74~MeV for Ca, whereas the distances between $1d_{3/2}$ and the Fermi level $\alpha_F$ are 3.5~MeV for Ar and 3.8~MeV for Ca. To reconstruct the missing data, we assume that all the neutron levels follow the same pattern, see Table~\ref{tab:ArEnergyLev}. Due to the lack of knowledge about the proton levels of argon, we use the modified values from calcium. The data for oxygen~\cite{ref:OxLev_Gillet&Vinh,ref:Bohr&Mottelson} are collected in Table~\ref{tab:OxEnergyLev}.

The widths for most of the calcium levels can be determined by fitting to the plots of energy distribution in papers~\cite{ref:CaPLev_Tornow&Chen&Delaroche,ref:CaNLev_Johnson&Mahaux}. We estimate the remaining ones using the fact that~$D_\alpha$ should be, approximately, a function of a distance from the Fermi level~\cite{ref:CaNLev_Johnson&Mahaux, ref:CaPLev_Tornow&Chen&Delaroche, ref:ArLev_Johnson&Carlton&Winters, ref:deWittHuberts}:
\[
D_\alpha\propto\frac{(E_\alpha-E_F)^2}{(E_\alpha-E_F)^2+a^2}~.
\]
To get~$D_\alpha$'s for argon, we assume that their values lie on roughly the same curve as for Ca.

We did not find the energy distribution for oxygen, calculated in the same way as for calcium in Refs.~\cite{ref:CaPLev_Tornow&Chen&Delaroche,ref:CaNLev_Johnson&Mahaux}. Since oxygen nucleus plays only the role of a~testing ground for our model, we decided to obtain the proton $D_\alpha$'s directly from the energy distribution in the Benhar SF, and use the same values for neutrons. Thus we avoided additional discrepancies between the two descriptions.

\section{Results}\label{sec:Results}

\subsection{Electron scattering}\label{sec:Electrons}%

The goal of this subsection is to confront the model presented in Sec.~\ref{sec:Description} with the existing electron scattering data. Since description of the dip region and the $\Delta$~excitation is ambiguous~\cite{ref:Benhar&Farina&Nakamura,ref:Gil&al.,ref:MAID}, our considerations include QE interactions only and we test predictions of the obtained SFs in energy transfers below the QE peak. Figures~\ref{fig:Oxy}--\ref{fig:Ar} show comparison with a~wide spectrum of experimental points. The missing cross section for energy transfer above the QE peak may be attributed to the two-nucleon interactions, $\Delta$ production, and nonresonant background. In the captions, we give the momentum transfer at the QE peak calculated according to the formula
\[
\n q=\sqrt{\omega^2+ 2E_{\ve k}(E_{\ve k}-\omega )(1-\cos\theta_e)}~.
\]
This value depends rather weakly on~$\omega$ and therefore it provides quite good characteristics of the whole peak.

\subsubsection{Oxygen}\label{sec:Oxygen}%

We start with the oxygen target. Figure~\ref{fig:Oxy} presents the predictions of three models and the data from Refs.~\cite{ref:Anghinolfi,ref:Anghinolfi_Ar,ref:O'Connell}. The dotted line corresponds to the Fermi gas (FG) model (Fermi momentum $p_F=225$~MeV, binding energy $\epsilon_B=25$~MeV, no FSI), the solid line shows the cross sections of the oxygen GSF with FSI as in Sec.~\ref{sec:FSI}, and the dashed line depicts results for the Benhar SF with the same FSI. Differences between our model and the more systematic SF are of the size of the error bars. The main source of these differences is another momentum distribution, see Fig.~\ref{fig:OxBenhar} and the discussion in Sec.~\ref{sec:Precision}. Both SFs reproduce the shape and height of the QE peak quite well, but underestimate the cross section at low~$\omega$'s. This discrepancy may be attributed to the unsatisfactory treatment of FSI effects, because they tend to increase the cross section in this region. The best agreement with the data is obtained for the 880-MeV electron beam, whereas the worst one corresponds to $E_{\ve k}=537$~MeV. Fortunately, the latter set of data is least relevant in our analysis (see Sec.~\ref{sec:Selection}).

\subsubsection{Calcium}\label{sec:Calcium}%

For calcium, we compare in Fig.~\ref{fig:Ca} the cross sections obtained using the FG model ($p_F=249$~MeV, $\epsilon_B=33$~MeV~\cite{ref:Whitney}; represented by the dotted line), the GSF (solid line), and the calculations of Butkevich and Mikheyev~\cite{ref:Butkevich&Mikheyev} (dashed line) to the sample of electron scattering data collected at scattering angle 45.5$^\circ$ and various beam energies~\cite{ref:Williamson}. Only our model includes FSI effects.

The FG model describes very well the position and size of the QE peak for the highest values of beam energy. When the energy is lower than 700~MeV, it obviously fails.

Despite the fact that the approach of Butkevich and Mikheyev~\cite{ref:Butkevich&Mikheyev} is based on the SF, it yields results very similar to these for the FG model when $\omega$ is near the value corresponding to the QE peak or higher. The reason of this behavior lies in too simple treatment of the MF part in their SF: the energy distribution was limited to a~single $\delta$~function.

For energies 628--841~MeV, the accuracy of the GSF is very good. The occurring discrepancies can be explained as a contribution from the~$\Delta$ production. At small values of energy transfer, the cross section is  slightly overestimated. It means also that the QE peak is slightly underestimated, because FSIs based on a~folding function do not change the total cross section. Note that the agreement with the data in our region of interest (see Sec.~\ref{sec:Selection}) is better than in the case of oxygen. It may be attributed to the way FSI is introduced: Density of nucleus is assumed to be constant and equal to the saturation density of nuclear matter; this approximation should work better for heavier nuclei. The real part of the optical potential used in our computations was obtained for calcium and should work better for this target than for oxygen.

For $E_{\ve k}\leq545$~MeV, our model fails to describe the position and shape of the QE peak. However, the inaccuracy of the FG and the approach of Ref.~\cite{ref:Butkevich&Mikheyev} is visibly more severe. Similar problems for oxygen occur when $E_{\ve k}\leq700$~MeV. At the first glance, there is no connection between these two cases.  But when we have a closer look at the values of the momentum transfer at the QE peak, we will discover that our model starts to lose accuracy when momentum transfer is lower than $\sim$350--400~MeV. It could be related to simplifying assumptions of our approach, treatment of FSI effects or the very basic assumption---the IA. The models~\cite{ref:Benhar&Farina&Nakamura,ref:FSI_Horikawa&al.} based on different from our approximations apart from the IA and with more systematical treatment of FSI suffer a~similar drawback. It suggests that it is the loss of reliability of the IA, what is responsible for the problem.

\subsubsection{Argon}\label{sec:Argon}%

Before applying our model to neutrino interactions, we perform the final test by confronting it with the data for electron scattering off argon. We have found only one such experiment~\cite{ref:Anghinolfi_Ar}, which measured the cross section of $700$-MeV electrons scattered at $32^\circ$.

In the left panel of Fig.~\ref{fig:Ar}, predictions of the argon GSF and the FG model ($p_F=251$ MeV and $\epsilon_B=28$ MeV) are presented. The accuracy of the GSF is clearly better than that of the FG model. The result for our model, shown by the solid line, does not describe properly only the cross section at very low values of energy transfer. We have faced the same problems for oxygen and calcium and interpreted it as a~breakdown of the IA at $\n q\alt350$--400~MeV. In the considered case of scattering off argon, the momentum transfer at the QE peak is equal to 371~MeV. When we compare the result for argon with the one for oxygen in exactly the same kinematical conditions (see right panel of Fig.~\ref{fig:Ar}), we can see that the level of accuracy is comparable. The same holds true also for comparison with scattering off calcium for electron-beam energy 471 or 545~MeV. Therefore, we expect that if the typical~$\n q$ was higher, the agreement with the data for argon would be better.

We have observed that even for argon, the neutron SF may be approximated by the corresponding proton SF, as far as electron scattering is concerned. It can be explained by the fact that the contribution of neutrons to the inclusive cross section is small, what suppresses the differences between the SFs. This contribution is equal to 13\% for 700-MeV electrons scattered at 32$^\circ$ and rises to 23\% when the beam energy is increased to 1200~MeV.

\subsection{Neutrino scattering}\label{sec:Neutrinos}%

\begin{figure}
        \includegraphics[width=0.46\textwidth]{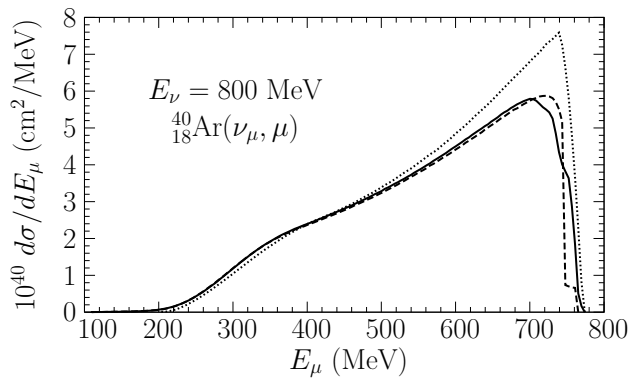}%
\caption{\label{fig:nuAr800} Quasielastic differential cross section
$d\sigma^\text{weak}/dE_\mu$ of \isotope[40][18]{Ar} as a~function of produced
muon energy $E_\mu$ for the GSF (solid line), approach of
Ref.~\cite{ref:Ankowski&Sobczyk}  (dashed line), and the FG model
(dotted line).}
\end{figure}

In the case of neutrino scattering, quantities of interest are the total cross section and the differential cross section in $Q^2=-q^2$ or in energy transfer (equivalently: in energy of produced muon).

Figure~\ref{fig:nuAr800} depicts differences between $d\sigma^\text{weak}/dE_\mu$ for the argon GSF (solid line), the
SF we described in Ref.~\cite{ref:Ankowski&Sobczyk} (dashed line), and the FG model (dotted line). One can see that the SFs introduce significant reduction of the cross section, mainly in the region of low energy transfers. The line representing the predictions of the GSF model is slightly wiggly, because when $\omega$ increases, lower-lying energy levels consecutively start contributing to the cross section. There are no singularities in the cross section, and in this sense, the GSF is more realistic then the SF from Ref.~\cite{ref:Ankowski&Sobczyk}. Effects of  FSI are not taken into account except those from Pauli blocking, but their influence on the cross section $d\sigma^\text{weak}/dE_\mu$ is rather small (see Fig.~14 in Ref.~\cite{ref:Benhar&Farina&Nakamura} showing the
impact of introducing FSI on $d\sigma^\text{weak}/dQ^2$ ). The purpose of Fig.~\ref{fig:nuAr800} is to show discrepancy of our description of argon nucleus and the FG model, commonly used in Monte Carlo simulations.

The results for neutrinos cannot be directly confronted with experimental data. Therefore, we first identified, in Sec.~\ref{sec:Selection}, the region in the $(\omega, \n q)$ plane which is most important for the 800-MeV neutrino scattering. Than we substantiated accuracy of our approach: we showed in Sec.~\ref{sec:Electrons} that it describes well kinematical aspects of nuclear effects. This whole analysis allows us to expect that using the presented approximation of the SF, we model neutrino interactions at a~similar level of accuracy as achieved in the case of electron scattering.

\section{Discussion of precision}\label{sec:Precision}%

Our approach is based on many approximations and in this section, we would like to understand how uncertain our final predictions are.

\begin{figure}
    \includegraphics[width=0.46\textwidth]{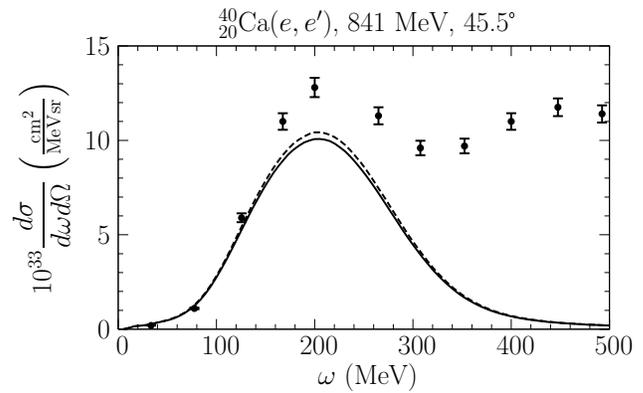}%
\caption{\label{fig:CaDip} Dependence of the cross section on the form factors. The dipole parametrization (dashed line) produces $\sim$3\% higher result than the BBBA05 one~\cite{ref:BBBA05}.}
\end{figure}

\begin{figure}
        \includegraphics[width=0.46\textwidth]{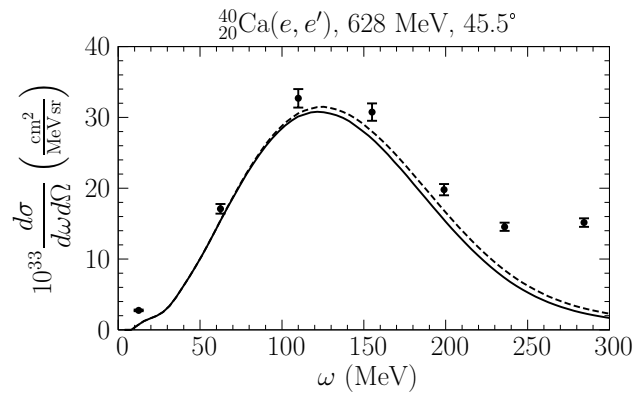}%
\caption{\label{fig:Ca628noCEC} Influence of the procedure restoring current conservation [Eq.~\eqref{eq:CECRestoration}] on the cross section. The result obtained without it is represented by the dashed line.}
\end{figure}

\begin{figure}
        \includegraphics[width=0.46\textwidth]{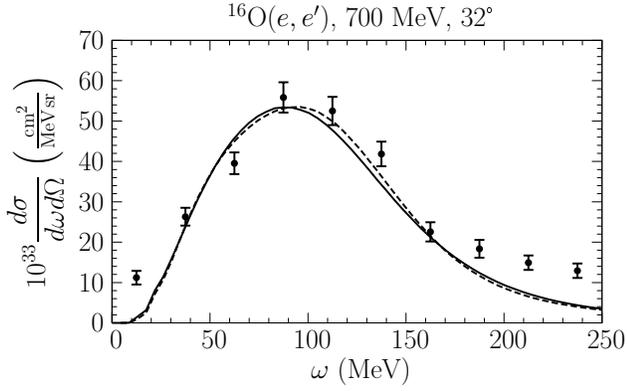}%
\caption{\label{fig:OxBenhar} Intrinsic inaccuracy of our model
arising from the treatment of the MF part of the SF.
Calculation for the Benhar's exact SF of oxygen (dashed line) are
compared with result for the GSF with the same momentum
distribution.}
\end{figure}

\subsubsection{General remarks}

\paragraph{Form factors.} Different choices of parameterization of the electromagnetic form factors may change the results by a~few percentages. As shown in Fig.~\ref{fig:CaDip}, the dipole parameterization yields the cross sections higher than the BBBA05 one~\cite{ref:BBBA05} used in this article. The discrepancy at the QE peak is $\sim$3.2\% for beam energy 350~MeV and $\sim$3.5\% for 841~MeV.

\paragraph{Current conservation.} Describing both electron and neutrino interactions, we applied the de Forest prescription to describe the off-shell kinematics. However, this leads to a~loss of conservation of the electromagnetic current in electron scattering and of the vector current in neutrino interactions. The procedure~\eqref{eq:CECRestoration}, by which we restore it in the electron case, modifies the cross section mainly above the QE peak; see Fig.~\ref{fig:Ca628noCEC}. For beam energy 628~MeV, the effect is as small as 1.5\% at the peak and it
decreases when the energy becomes larger.

\paragraph{Simplifications in the mean-field SF.} In the derivation of Eq.~\eqref{eq:PMF}, we made two simplifying assumptions: level widths do not depend on momentum and contribution to the momentum distribution of each level is the same [see Eq.~\eqref{eq:eachLevSameMD}]. Figure~\ref{fig:OxBenhar} illustrates the loss of accuracy due to these simplifications. To depict their influence, we use the momentum distribution calculated from the Benhar SF instead of the one from Ref.~\cite{ref:Bisconti&Arias&Co}. Since the level widths of oxygen are obtained by fitting to the energy distribution of the Benhar SF, a~slightly different shape of the predicted QE peak is the result of the simplifying assumptions only. We checked that for other values of beam energy discrepancy does not increase. Therefore, we conclude that the GSF can be considered as quite good approximation of the more systematic approach.

\paragraph{Parameterization of the momentum distributions.} Application of the momentum distributions from Ref.~\cite{ref:Bisconti&Arias&Co} in our model requires dividing each of them into the MF and correlated parts. It involved introduction of a~few parameters. To find out how much choice of these parameters influences the cross sections, we calculated first \nC~for the oxygen normalized as in Table~\ref{tab:MomDistrib}, but with $e_1=1.770$ (instead of 1.400):
\begin{eqnarray*}
\nC(\ve p)&=&\frac{1.02}{(2\pi)^3}\frac{16}{8}\big[2.670\exp(-1.770\:\ve p^2)\\
& &\qquad\qquad+\:0.2128\exp(-0.303\:\ve p^2)\big]
\end{eqnarray*}
for $0\leq\n p\leq2.025$~fm$^{-1}$. When it is applied, the cross sections change less than 0.1\% in the considered energy range. Thus, we do not need to pay much attention to parameter $e_1$, as far as the same normalization is kept. Second, we found the distribution with $e_1=1.400$, but with the normalization 16.2\% (instead of 12.0\%):
\begin{eqnarray*}
\nC(\ve p)&=&\frac{1.02}{(2\pi)^3}\frac{16}{8}\big[3.9228\exp(-1.400\:\ve p^2)\\
& &\qquad\qquad+\:0.0736\exp(-0.091\:\ve p^2)\big]
\end{eqnarray*}
at the interval $0\leq\n p\leq2.025$~fm$^{-1}$. The above distribution leads to the cross sections changed by up to 2.2\%. We have analyzed a~few such modifications and in each case we have found that the influence of the normalization is greater than that of $e_1$. It is because variation of the parameter $e_1$ only redistributes the strength within given part of the momentum distribution [and as a~consequence modifies the parameter $\alpha$ in Eq.~\eqref{eq:correlationSF}], whereas variation of the normalization changes the way some part of the strength is treated.

\begin{figure}
        \includegraphics[width=0.46\textwidth]{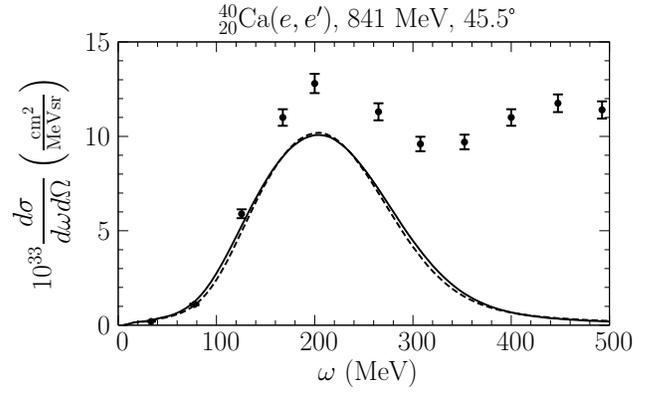}%
\caption{\label{fig:CaCdA} Uncertainty of the the cross section
with respect to used momentum distribution. Solid line shows
result for momentum distribution from
Ref.~\cite{ref:Bisconti&Arias&Co} (used throughout this paper) and the
dashed one from Ref.~\cite{ref:Ciofi&Simula}.}
\end{figure}

\begin{figure}
        \includegraphics[width=0.46\textwidth]{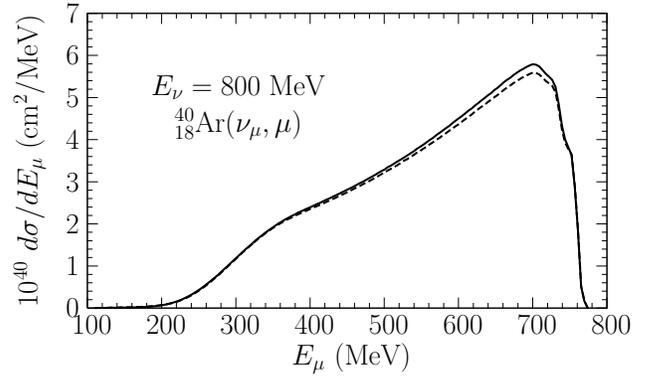}%
\caption{\label{fig:ArCdA} Same as Fig.~\ref{fig:CaCdA} but for
$\nu_\mu$ scattering off argon.}
\end{figure}

\paragraph{Momentum distributions.} Both for oxygen and calcium, the momentum distributions are given by analytical formulas in Ref.~\cite{ref:Ciofi&Simula}. In Fig.~\ref{fig:CaCdA}, we show that even though they predict slightly higher QE peak, the yielded cross section is lower. Because the calcium momentum distributions are used for argon, its description ``inherits'' the same uncertainties, see Fig.~\ref{fig:ArCdA}. Throughout this paper, we rely on the distributions from Ref.~\cite{ref:Bisconti&Arias&Co}, because they are obtained in more systematic calculations.

\begin{figure}
        \includegraphics[width=0.46\textwidth]{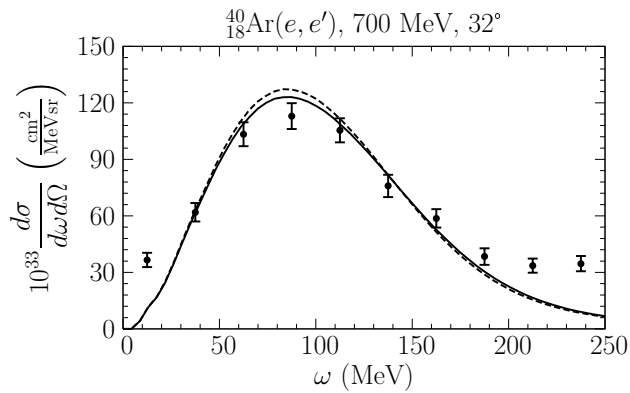}%
\caption{\label{fig:Ca48toAr} Estimation of the uncertainty due
to unknown momentum distribution of argon. The cross section
calculated using the SF of \isotope[40][18]{Ar} with momentum
distribution of \isotope[40][20]{Ca} (solid line) and
\isotope[48][20]{Ca} (dashed line).}
\end{figure}

\begin{figure}
        \includegraphics[width=0.46\textwidth]{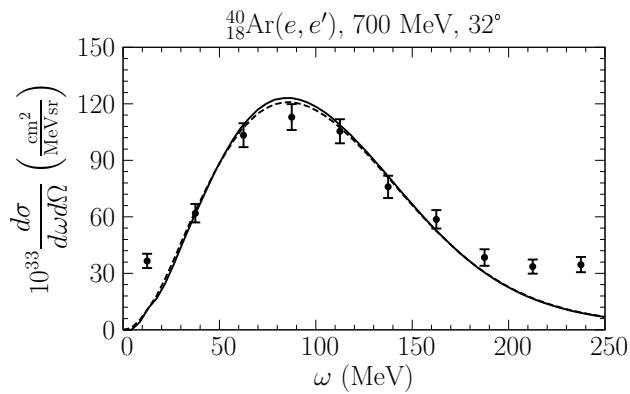}%
\caption{\label{fig:ArSpreading} Influence of the level width
on the cross section. Solid line: calculation with the values from
Table~\ref{tab:ArEnergyLev}. Dashed line: values multiplied by~3.}
\end{figure}

\begin{figure}
        \includegraphics[width=0.46\textwidth]{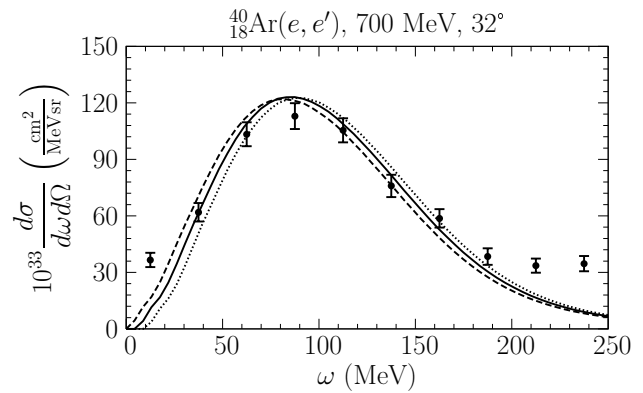}%
\caption{\label{fig:ArLev} Comparison of the cross sections
obtained with the energy levels from Table~\ref{tab:ArEnergyLev}
(solid line), levels shifted by $+5$~MeV (dotted line), and by
$-5$~MeV (dashed line).}
\end{figure}

\subsubsection{Case of argon}

In addition to the already described sources of uncertainty, the description of argon nucleus suffers from the lack of the available momentum distributions and knowledge of energy levels. We estimate them using the information for \isotope[40][20]{Ca}. For this reason, a few words of comment on the accuracy for this specific nucleus are needed.

\paragraph{Momentum distributions.} The surplus neutrons modify both the proton and neutron momentum distributions. A~similar situation appears for \isotope[48][20]{Ca}, where the distributions are available~\cite{ref:Bisconti&Arias&Co}. We have used the \isotope[48][20]{Ca} momentum distributions to estimate how these modifications can affect the argon cross sections; see Fig.~\ref{fig:Ca48toAr}. The proton cross section was increased by 4\% and the neutron one was decreased by 3.8\%. The overall increase is equal to 2.9\%. The number of surplus neutrons in \isotope[40][18]{Ar} is smaller than in \isotope[48][20]{Ca}, therefore we expect this effect to be smaller too.

\paragraph{Level widths.} Due to the lack of any knowledge about the level widths of argon, we use the values for calcium. Figure~\ref{fig:ArSpreading} presents that~$D_\alpha$'s three times larger than those given in Table~\ref{tab:ArEnergyLev} change the cross sections only up to 2\% (decrease at the peak). Narrower levels gives barely noticeable difference: 0.23\% for the widths divided by~3, and 0.53\% for divided by 100 (increase at the peak). For $d\sigma^\text{weak}/dE_\mu$, the more the levels overlap, the less wiggly the cross section is.

\paragraph{Energy levels.} The argon energy levels may differ from the used ones. We may expect that the discrepancies in Table~\ref{tab:ArEnergyLev} are distributed randomly, and so a~part of their influence on the cross section is diminished. Figure~\ref{fig:ArLev} shows that even if every level is shifted by the same value, chosen to be 5~MeV, the cross section does not change dramatically---the QE peak only moves a little bit. We conclude that the way to increase the accuracy of the presented argon SF is to apply the actual values of the energy levels; the degree in which they are smeared has minor influence on the cross section, especially in the case of electrons.

\subsubsection{Final state interactions}

\paragraph{Real potential.} To find out if one can approximate the real part of the potential by a~constant, we have applied the value 10~MeV. In the case of oxygen, it slightly improved agreement with the experimental data. However, the same value employed to calcium decreased the level of accuracy of the model. It might suggest that the real potential for oxygen is deeper than the potential shown in Fig.~\ref{fig:pot}.

\paragraph{Imaginary potential.} The use of the imaginary part of the potential $U(\ve{p'})$ defined in Eq.~\eqref{eq:OP} instead of approximation~\eqref{eq:imaginaryOP} has minor influence on the obtained cross sections. Typical change is a~$\sim$1\%-increase. We conclude that for practical purposes these two approaches are equivalent.

\paragraph{Cross section.} When evaluating imaginary potential~\eqref{eq:imaginaryOP}, we have fixed the nucleon-nucleon cross section to 17.4~mb, which corresponds to nucleon kinetic energy 200~MeV. In principle, one should take into account the cross section's dependence on energy. Therefore, to check validity of our approximation, we have used the exact nucleon-nucleon cross section~\cite{ref:Pandharipande&Pieper} in the energy range 100--300 MeV, most important for the discussed kinematical region. For 545-MeV electron scattering off calcium, the result decreases by 1.1\%. When beam energy is higher, the effect is even smaller.

\paragraph{Density of nucleus.} We have assumed that the density of nucleus is equal to the saturation density,
despite the fact that in reality its average value is smaller. However, the quantity of interest is not the density
itself but rather $\rho_\text{nucl}\sigma_{N\!N}$. This product decreases by 7\% (15.4\%) when $\rho_\text{nucl}$ changes to 0.14~fm$^{-3}$ (0.12 fm$^{-3}$), i.e., by 12.5\% (25\%). Since the corresponding increase of the electron cross section is only 1\% (2.4\%) at the QE peak, our approach seems to be well justified.

\paragraph{Folding function.} Employing Lorentzian folding function, i.e., neglecting correlations between nucleons in nucleus is a~crude approximation~\cite{ref:FSI_Benhar&al, ref:Benhar&Day&Sick,ref:Benhar_transparency}. Comparison to the results presented in Ref.~\cite{ref:Benhar&Farina&Nakamura} suggests that an accurate approach could yield the cross sections higher at the peak by up to  $\sim$15\% and with lower tails. Precise comparison is difficult because in our computations contribution of the $\Delta$ resonance is missing.

\section{Summary}\label{sec:Summary}%

The main goal of the article is to improve description of neutrino scattering off argon in the 1-GeV energy region. We have presented the way to calculate approximate spectral functions of medium nuclei and applied it to electron scattering off oxygen, calcium, and argon targets. For neutrino interactions precise experimental data are missing. Therefore, we have identified the region of the $(\omega, \n q)$ plane which is most important for neutrino
quasielastic interaction. The presented model to describe nuclear effects has been then tested using the electron scattering data which lie in this region. The obtained agreement is good in the case of oxygen and very good for calcium. Moreover, our approximation reproduces results of the Benhar SF for oxygen with a~satisfactory degree of accuracy. Detailed discussion of uncertainties due to many simplifications of our model have lead us to the conclusion that all of them are of the order of a~few percentages.

In addition, we have observed that when the typical value of the momentum transfer is less than $\sim$350--400~MeV, systematic discrepancies between the presented model and the electron data occur: the shape of the calculated cross section $\frac{d\sigma}{d\omega d\Omega}$ is not suitable to fit the data and also increasing amount of strength is missing at low energy transfers. A~similar problem is present in other models~\cite{ref:Benhar&Farina&Nakamura,ref:FSI_Horikawa&al.} what suggests that the source of the problem may be the loss of reliability of the impulse approximation.

In this paper, we tried to give all the ingredients used in our numerical computations to allow implementation of our spectral functions in neutrino Monte Carlo generators.

\begin{acknowledgments}
We would like to thank Giampaolo Co' for the momentum distributions used in this paper. We also express our gratitude to Omar Benhar for providing us with his spectral function for oxygen. The authors were supported by MNiSW under Grants No. 3735/H03/2006/31 (JTS, AMA) and No. 3951/B/H03/2007/33 (AMA).
\end{acknowledgments}


\begin{thebibliography}{99}
%
\bibitem{ref:NuInt}%
    {J.~G. Morfin, M.~Sakuda, and Y.~Suzuki (eds.), Nucl. Phys. B, Proc. Suppl. {\bf 112} (2002); %
    F.~Cavanna, P.~Lipari, C.~Keppel, and  M.~Sakuda (eds.) {\it ibid} {\bf 139} (2005); %
    F.~Cavanna, J.~G. Morfin, and T.~Nakaya (eds.) {\it ibid} {\bf 159} (2006).}%
%
\bibitem{ref:NextGen}
    {A.~Blondel, Acta Phys. Pol.~B {\bf37}, 2077 (2006);
    A.~Ereditato and A.~Rubbia, Nucl. Phys. B, Proc. Suppl. {\bf 154}, 163 (2006);
    G.~Battistoni, A.~Ferrari, C.~Rubbia, P.~R.~Sala, and F.~Vissani, arXiv:hep-ph/0604182.}%
%
\bibitem{ref:Ankowski&Sobczyk}%
    {A.~M.~Ankowski and J.~T.~Sobczyk, Phys. Rev. C {\bf 74}, 054316 (2006).}%
%
\bibitem{ref:Ankowski}%
    {A.~M.~Ankowski, Acta Phys. Pol.~B {\bf37}, 2259 (2006).}%
%
\bibitem{ref:Benhar&Farina&Nakamura}%
    {O.~Benhar, N.~Farina, H.~Nakamura, M.~Sakuda, and R.~Seki, Phys. Rev. D {\bf 72}, 053005 (2005).}%
%
\bibitem{ref:Butkevich&Mikheyev}%
    {A.~V.~Butkevich and S.~P.~Mikheyev, Phys. Rev. C {\bf 72}, 025501 (2005).}%
%
\bibitem{ref:Anghinolfi_Ar}%
    {M.~Anghinolfi {\it et al.}, J. Phys. G 21, L9 (1995).}%
%
\bibitem{ref:Frullani&Mougey}%
    {S.~Frullani and J.~Mougey, Adv. Nucl. Phys. {\bf 14}, 1 (1984).}
%
\bibitem{ref:BBBA05}%
    {R.~Bradford, A.~Bodek, H.~Budd, and J.~Arrington, Nucl. Phys. B, Proc. Suppl. {\bf 159}, 127 (2006).}%
%
\bibitem{ref:deForest}%
    {T. de Forest Jr., Nucl. Phys. {\bf A392}, 232 (1983).}%
%
\bibitem{ref:Maieron&al}%
    {C.~Maieron,  M.~C. Mart\'inez, J.~A. Caballero, and J.~M. Ud\'ias, Phys. Rev. C {\bf 68}, 048501 (2003).}%
%
\bibitem{ref:FSI_Benhar&al}%
    {O.~Benhar, A.~Fabrocini, S.~Fantoni, G.~A.~Miller, V.~R.~Pandharipande, and I.~Sick, Phys. Rev. C {\bf 44}, 2328 (1991).}%
%
\bibitem{ref:FSI_Nakamura&Seki&Sakuda}%
    {H.~Nakamura, R.~Seki, and  M.~Sakuda, Nucl. Phys. B, Proc. Suppl. {\bf 139}, 201 (2005).}%
%
\bibitem{ref:FSI_Co'}%
    {G.~Co', Nucl. Phys. B, Proc. Suppl. {\bf 159}, 192 (2006).}%
%
\bibitem{ref:Oset&al_nonres}%
    {E.~Oset, L.~L.~Salcedo, D.~Strottman, Phys. Lett. {\bf B165}, 13 (1985).}%
%
\bibitem{ref:Oset&Salcedo_nonres}%
    {E.~Oset, L.~L.~Salcedo, Nucl. Phys. {\bf A468}, 631 (1987).}%
%
\bibitem{ref:Gross&Lipperheide}%
    {D.~H.~E. Gross and R.~Lipperheide, Nucl. Phys. {\bf A150}, 449 (1970).}%
%
\bibitem{ref:Anghinolfi}%
    {M.~Anghinolfi {\it et al.}, Nucl. Phys. {\bf A602}, 405 (1996).}%
%
\bibitem{ref:O'Connell}%
    {J.~S.~O'Connell {\it et al.}, Phys. Rev. C {\bf 35}, 1063 (1987).}%
%
\bibitem{ref:Whitney}%
    {R.~R.~Whitney, I.~Sick, J.~R. Ficenec, R.~D. Kephart, and W.~P. Trower, Phys. Rev. C \textbf{9}, 2230 (1974).}%
%
\bibitem{ref:Meziani}%
    {Z.~E.~Meziani {\it et al.}, Phys. Rev. Lett. {\bf 54}, 1233 (1985).}%
%
\bibitem{ref:Yates}%
    {T.~C.~Yates {\it et al.}, Phys. Lett. {\bf B312}, 382 (1993).}%
    %
\bibitem{ref:Williamson}%
    {C.~F.~Williamson {\it et al.}, Phys. Rev. C {\bf 56}, 3152 (1997).}%
%
\bibitem{ref:Ciofi&Simula&Frankfurt&Strikman}%
    {C.~Ciofi degli Atti, S.~Simula, L.~L. Frankfurt, and M.~I. Strikman, Phys. Rev. C {\bf 44}, R7 (1991).}
%
\bibitem{ref:Kulagin&Petti}%
    {S.~A.~Kulagin and R.~Petti, Nucl. Phys. {\bf A765}, 126 (2006).}%
%
\bibitem{ref:Benhar&Fabrocini&Fantoni&Sick}%
    {O.~Benhar, A.~Fabrocini, S.~Fantoni, and I.~Sick, Nucl. Phys. {\bf A579}, 493 (1994).}%
%
\bibitem{ref:Ciofi&Liuti&Simula}%
    {C.~Ciofi degli Atti, S.~Liuti, and S.~Simula, Phys. Rev. C {\bf 41}, R2474 (1990).}%
%
\bibitem{ref:FSI_Horikawa&al.}%
    {Y.~Horikawa, F.~Lenz, and N.~C.~Mukhopadhyay, Phys. Rev. C {\bf 22}, 1680 (1980).}%
%
\bibitem{ref:Pandharipande&Pieper}%
    {V.~R.~Pandharipande and S.~C.~Pieper, Phys. Rev. C {\bf 45}, 791 (1992).}%
%
\bibitem{ref:Cooper&al}%
    {E.~D. Cooper, B.~C. Clark, R.~Kozack, S.~Shim, S.~Hama, J.~I. Johansson, H.~S. Sherif, R.~L. Mercer, and D.~B. Serot, Phys. Rev. C {\bf 36}, 2170 (1987).}%
%
\bibitem{ref:Cooper&Hama&Clark&Mercer}%
    {E.~D. Cooper, S.~Hama, B.~C. Clark, and R.~L. Mercer, Phys. Rev. C {\bf 47}, 297 (1993).}%
%
\bibitem{ref:Bisconti&Arias&Co}%
    {C.~Bisconti, F.~Arias de Saavedra, and G.~Co', Phys. Rev. C {\bf 75}, 054302 (2007).}%
%
\bibitem{ref:Co'_priv}%
    {G.~Co', private communication.}
%
\bibitem{ref:MD_Benhar&al}%
    {O.~Benhar, C.~Ciofi degli Atti, S.~Liuti, and G. Salm\`e, Phys. Lett. {\bf B177}, 135 (1986).}%
%
\bibitem{ref:Ciofi&Simula}%
    {C.~Ciofi degli Atti and S.~Simula, Phys. Rev. C {\bf 53}, 1689 (1996).}%
%
\bibitem{ref:CaNLev_Johnson&Mahaux}%
    {C.~H.~Johnson and C.~Mahaux, Phys. Rev. C {\bf 38}, 2589 (1988).}%
%
\bibitem{ref:CaPLev_Tornow&Chen&Delaroche}%
    {W.~Tornow, Z.~P.~Chen, and J.~P.~Delaroche, Phys. Rev. C {\bf 42}, 693 (1990).}%
%
\bibitem{ref:ArLev_Johnson&Carlton&Winters}%
    {C.~H.~Johnson, R.~F.~Carlton, and R.~R.~Winters, Phys. Rev. C {\bf 44}, 657 (1991).}%
%
\bibitem{ref:OxLev_Gillet&Vinh}%
    {V.~Gillet and N.~Vinh Mau, Nucl. Phys. {\bf 54}, 321 (1964).}%
%
\bibitem{ref:Bohr&Mottelson}%
    {A.~Bohr and B.~R. Mottelson, {\it Nuclear Structure} (Benjamin, New York, 1969), Vol.~1, pp.~232,~326.}%
%
\bibitem{ref:deWittHuberts}%
    {P.~K.~A. de Witt Huberts, J. Phys. G 16, 507 (1990).}%
%
\bibitem{ref:Gil&al.}%
    {A.~Gil, J.~Nieves, and E.~Oset, Nucl. Phys. {\bf A627}, 543 (1997).}%
%
\bibitem{ref:MAID}%
    {D.~Drechsel, O.~Hanstein, S.~S. Kamalov, and L.~Tiator, Nucl. Phys. {\bf A645}, 145 (1999).}%
%
\bibitem{ref:Benhar&Day&Sick}%
    {O.~Benhar, D.~Day, and I.~Sick, Rev. Mod. Phys. {\bf 80}, 189 (2008).}%
%
\bibitem{ref:Benhar_transparency}%
    {O.~Benhar, Nucl. Phys. B, Proc. Suppl. {\bf 159}, 168 (2006).}%
\end{thebibliography}
\end{document}